\documentclass[11pt]{article}

\usepackage[utf8]{inputenc}
\usepackage[T1]{fontenc}
\usepackage[english]{babel}

\usepackage{amsmath, amssymb, amsthm}
\usepackage{mathrsfs}
\usepackage{stmaryrd}
\usepackage{verbatim}
\usepackage{enumerate}

\usepackage{geometry}
\usepackage{hyperref}
\hypersetup{
  colorlinks=true,
  linkcolor=blue,
  citecolor=blue,
  urlcolor=blue
}

\usepackage{graphicx}
\usepackage{enumitem}
\usepackage{booktabs}
\usepackage{xcolor}

\usepackage{cite}

\usepackage{authblk}
\newtheorem{theorem}{Theorem}[section]
\newtheorem{lemma}[theorem]{Lemma}
\newtheorem{definition}[theorem]{Definition}
\newtheorem{proposition}[theorem]{Proposition}

\theoremstyle{remark}

\title{From Knowledge to Conjectures: \\
A Modal Framework for Reasoning about Hypotheses}
\author{Fabio Vitali\thanks{fabio.vitali@unibo.it}}
\affil{University of Bologna, Italy}

\begin{document}
\maketitle

\begin{abstract}
This paper introduces a new family of cognitive modal logics designed to formalize conjectural reasoning: modal systems in which cognitive contexts extend known facts with hypothetical assumptions in order to explore their consequences. Unlike traditional doxastic and epistemic systems, conjectural logics rely on a principle, called Axiom \textbf{C} ($\varphi \rightarrow \Box\varphi$), through which established facts are preserved across conjectural layers. While Axiom \textbf{C} has often been treated with suspicion because of its association with modal collapse, we show that collapse does not arise from \textbf{C} alone, but requires either the presence of Axiom \textbf{T} or a concretely bivalent base logic. Accordingly, we avoid \textbf{T} and adopt a non-bivalent semantic framework, such as supervaluation-style semantics, Weak Kleene logic, or Description Logic, in which undefined propositions may coexist with modal assertions. This prevents modal collapse and preserves a distinction between factual and conjectural statements. Within this framework we define the modal systems $\mathbf{KC}$ and $\mathbf{KDC}$, show that Axiom \textbf{C} directly implies \textbf{4} and \textbf{5}, and prove that these systems are non-trivial, sound, and complete. An inclusion theorem links reality, doxastic states, epistemic states, and conjectural states via set-theoretic inclusion among valuations, providing a unified account of how these layers relate. Finally, we introduce a dynamic operator, $\mathsf{settle}(p)$, which formalizes the transition by which a conjectural extension becomes designated reality, thereby motivating a corresponding Conjectural Dynamic Logic.
\end{abstract}

\begin{flushleft}
\end{flushleft}

\section{Introduction}\label{sec:intro}
We call \textit{cognitive modal logics} a family of systems designed to model subjective informational states, i.e., agents' beliefs, knowledge, or conjectures, rather than absolute necessity/possibility, obligation/permission, or temporal features. 
The two best-known families of cognitive logics are: \textit{doxastic logics}, that formalize belief states, i.e., possibly incorrect propositions held by an agent, where very little is assumed to hold and even internal consistency is an optional condition; and \textit{epistemic logics}, that formalize (partial) knowledge, i.e., propositions that are believed and also true: agents know a subset of all truths, and epistemic logics represent accurate but partial access to reality.

Cognitive modal logics are peculiar for some structural characteristics: first, their semantics explicitly allow for a separation between facts and subjective states. Second, they admit and represent multiple independent modal systems, i.e., one per agent or cognitive stance, each governed by its own relations and valuations. Third, they are typically centered on a main operator (i.e., $\Box$, or \textbf{K}, or \textbf{B}) to capture formulas cognitively associated to a subjective perspective, while the secondary operator (i.e., $\Diamond$) plays a minimal or negligible role. Finally, rather than treating the addition of modal axioms as increasing refinements of a single system, cognitive logics treat axioms, and combinations thereof, as the foundation of qualitatively distinct reasoning modes.

We believe these two families do not exhaust the relevant approaches to cognitive logics, and there is room for additional varieties. This paper introduces a third category: \textit{conjectural logics}. These are designed to model hypothetical reasoning based on true information: agents have access to all currently known facts, and adopt in addition one or more unverified assumptions in order to explore their possible consequences. More precisely, conjectures are considered as local and controlled extensions of an incomplete factual base: they preserve what is already settled, and determine only what, at the current stage, remains open.

Formally, conjectural logics are governed by a single modal axiom stipulating that all established predicates are adopted within the hypothetical scenario, in addition to non-established ones that are assumed as well. From this axiom, new logics can be drawn in which conjectures supplement facts, cognitive layers are constructed modally, and each new assumption defines a distinct cognitive perspective. 

This paper introduces therefore a family of modal logics based on an axiom, referred here as Axiom \textbf{C}, which takes the form $ \varphi \rightarrow \Box \varphi $. Axiom \textbf{C} is actually well established in the literature. It has been considered at least since the mid-20th century in the context of alethic logic, most notably in Kurt Gödel’s notes on the ontological argument, where necessary existence of certain positive properties is postulated to follow from their actual existence (cf. \cite{godel1970ontological}; \cite{benzmueller2025scott}).

Despite its early appearance, Axiom \textbf{C} has traditionally been considered with some caution due to its association with the phenomenon known as \textit{modal collapse}. Modal collapse occurs when modal distinctions are lost. Mathematically, this means that the modal operator $ \Box $ becomes indistinguishable from the identity operator, so that, for every formula $ \varphi $, we have $ \varphi \leftrightarrow \Box \varphi $: the distinction between modal and non-modal truths collapses, and the modal operator becomes logically trivial. 
Within an alethic perspective, modal collapse implies that all true propositions are \textit{necessarily true}: every actual fact becomes metaphysically necessary, leaving no room for contingency or possibility. This has deep implications in metaphysics and theology, as it would eliminate the conceptual space in which possible alternative states of reality could meaningfully be imagined.

However, we observe that Axiom \textbf{C} alone does not bring about modal collapse. Rather, the collapse arises from the conjunction of Axiom \textbf{C} and Axiom \textbf{T}, or from the assumption that the base logic is concretely bivalent. In a bivalent setting, the evaluation of all well-formed formulas, modal and non-modal alike, is governed by a total valuation assigning exactly one of two values, true or false, to each formula. Under the assumption of bivalence, or the simultaneous adoption of Axiom \textbf{T}, Axiom \textbf{C} cannot escape the equivalence $ \varphi \leftrightarrow \Box \varphi $, making modal formulas indistinguishable from their non-modal counterparts and eliminating any asymmetry between levels. Indeed, we claim that concrete non-bivalence is both necessary and sufficient, in the absence of \textbf{T}, to avoid the collapse.

On the other hand, non-bivalent semantics, such as supervaluation-based frameworks or Weak Kleene logic, allow well-formed formulas to remain neither true nor false. In such settings, a formula $ \varphi $ may be undetermined at the base world, while the modal formula $ \Box \varphi $ is determined and is still fully evaluable through its behavior across accessible worlds. For this reason, the equivalence $ \varphi \leftrightarrow \Box \varphi $, i.e., modal collapse, does not in general hold. The modal and non-modal layers remain logically distinct, and Axiom \textbf{C} can therefore be used without trivializing the modal operator.

The possibility of well-formed yet undefined statements can also be found in description logics when adopting an open-world assumption (OWA): the absence of information about a formula $ \varphi $ does not justify the conclusion that $ \varphi $ is false, but instead leaves its truth value undetermined. In this framework, knowledge bases are considered incomplete by default, and reasoning proceeds with the understanding that additional information may yet be discovered or inferred. For instance, given the axioms $ \mathsf{Painter} \sqsubseteq \mathsf{Person} $, $ \mathsf{Painter}(\mathsf{Leonardo}) $, and $ \mathsf{Person}(\mathsf{John}) $, the proposition $ \mathsf{Painter}(\mathsf{John}) $ is neither entailed nor refuted: it is simply unknown under the current state of information. In these frameworks, modal assertions such as $ \Box \varphi $ need neither track nor presuppose any precise truth value of $ \varphi $ in the base world. Instead, modal statements are evaluated with respect to their behavior in accessible interpretations or cognitive states, allowing the logic to maintain a distinction between the modal and non-modal levels.

The semantics of the modal operator also calls for a closer examination. In standard alethic modal logic, axiom \textbf{T} is central to the interpretation of $ \Box $ as ``necessity,'' as it ensures that all necessary truths are actually true. But in the absence of \textbf{T}, and within a logic where some propositions may be without a truth value, any alethic interpretation becomes conceptually shaky. In this paper, we focus on cognitive modal logics instead. Under this interpretation, Axiom \textbf{C} enables the construction of modal levels in which a base layer of established facts is augmented by a higher layer of conjectures: well-formed formulas that are neither true nor false in the base logic, but which are assumed as true under the modal operator. The basic systems generated by conjectural reasoning are called \textbf{KC} and \textbf{KDC}. We show that these systems are non-trivial, and that they are sound and complete.

These logics characterize conjectures as local extensions of an incomplete factual base that is neither discussed, denied, nor revised. This makes conjecture suitable for representing phenomena such as \textit{future contingents}, uncertain \textit{historical capta}, \textit{personal opinions}, \textit{disagreement}, and, more generally, \textit{uncertainty} which are traditionally difficult to faithfully represent in logical systems. The same framework also supports a generalized inclusion result relating doxastic, epistemic, and conjectural worlds related to some designated factual base. This result supplies a common set-theoretic backbone for the family of cognitive modal logics developed here.

Finally, we extend the framework dynamically by introducing a new operator, $\mathsf{settle}(p) $, that transforms a conjectural extension into the new designated reality. This leads to a corresponding dynamic logic, which, by symmetry with dynamic doxastic and dynamic epistemic logics, we call \textit{Conjectural Dynamic Logic} (\textit{CDL}). This dynamic view captures transitions by which what was previously only conjectured becomes factual within the cognitive space.

As mentioned, the framework proposed through logics such as \textbf{KC} and \textbf{KDC} is compatible with several non-bivalent semantic settings. In particular, the underlying base logic may be given by supervaluation-style semantics, Weak Kleene semantics \cite{Kleene1952Metamathematics}, or description-logical settings under open-world assumptions. At the same time, despite some surface similarities, the phenomena at issue here are not adequately captured by existing formalisms such as default logic, autoepistemic logic, ceteris paribus logics, or standpoint-based approaches. Axiom \textbf{C} generates a logic in which conjectural worlds are constrained by facts, locally determinative, and structurally related to a designated incomplete reality.

The paper is organized as follows. Section 2 discusses modal collapse and the recent history of Axiom C. Section 3 characterizes conjectures philosophically and semantically as local extensions of incomplete factual bases. Section 4 introduces \textbf{KC} and \textbf{KDC} and proves their main formal properties, in particular soundness, completeness, and the necessary and sufficient condition for avoiding collapse in systems based on Axiom \textbf{C}. Section 5 develops the set-theoretic results connecting cognitive modal logics to a designated factual base. Section 6 shows how non-bivalent base logics such as supervaluation semantics, Weak Kleene logic, and description logics can serve as bases for conjectural logics. Section 7 introduces the dynamic operator of settlement and the corresponding conjectural dynamic logic. Section 8 discusses apparently similar frameworks, such as default logic, autoepistemic logic, AGM belief revision, ceteris paribus logic, and standpoint logic, and shows that they address problems different from the ones considered here. Section 9 offers an alethic, rather than cognitive, interpretation of Axiom C and the related logics. Section 10 concludes with some reflections on the broader significance of the framework for reasoning under uncertainty and disagreement.

\section{A Brief History of Axiom \textbf{C} and the Modal Collapse} \label{sec:history}

The schema $\varphi \rightarrow \Box \varphi$ (which we call Axiom \textbf{C}) has long been regarded as both powerful and problematic. In classical modal logics, the combination of Axiom $\mathbf{C}$ with Axiom $\mathbf{T}$ ($\Box \varphi \rightarrow \varphi$) entails $\varphi \leftrightarrow \Box \varphi$,
the so-called \textit{modal collapse}. 

Gödel’s c.~1970 manuscript on the ontological argument \cite{godel1970ontological} develops axioms of positivity, essence, and necessary existence that jointly entail the existence of a necessary being, even though not explicitly discussing \textbf{C}. Oppy \cite{oppy1995ontological} offers an early critical discussion of Gödel’s argument and its variants, noting modal collapse among the problematic consequences. Anderson \cite{anderson1996} then examines, under classical assumptions, whether Gödel’s axioms yield collapse. Sobel \cite{sobel2004logic} establishes in detail that, in Gödel’s system, collapse is unavoidable: $ \varphi \leftrightarrow \Box \varphi $, so every true formula becomes necessarily true: the collapse trivializes the modal landscape and undermines the purpose of distinguishing between possible worlds. 

Kovač \cite{kovac2012modal} investigates Gödel’s proof in detail, providing explicit proofs that modal collapse ($\forall X,\forall x,(Xx \leftrightarrow \Box Xx)$) is inherent in the model, and shows that necessity of identity follows, i.e., $(a = b) \rightarrow \Box(a=b)$. Kovač also argues, on textual grounds, that Gödel likely intended the collapse as a consequence of his system.

In provability logic, $\Box$ receives a different reading. Solovay’s completeness theorem \cite{solovay1976} shows that in the interpretation where $\Box \varphi$ means "$\varphi$ is provable in arithmetic,” adopting $\varphi \rightarrow \Box \varphi$
would assert that every true statement is provably true. This would collapse the gap between truth and provability, so the axiom is excluded from the system.

Some authors modify the modal system itself to control or block the collapse. For instance, Marcos \cite{marcos2005paranormal} shows that many standard normal modal logics become “paranormal” by defining negation out of modal primitives, namely either $\neg \varphi$ means $\neg \Box \varphi$ (\textit{negation as non-necessity}) or $\neg \varphi$ means $ \Box \neg \varphi$ (\textit{negation as impossibility}). The first yields \textit{paraconsistency} (which means that inconsistencies, i.e., $\varphi \wedge \neg \varphi$, do not make every statement true); the second yields \textit{paracompleteness} (which means that there may be truth-value gaps, i.e., well-formed formulas that are neither true nor false). These principles yield settings where classical implications fail. Indeed, the usual passage from $\mathbf{C}$ ($\varphi \to \Box\varphi$) and $\mathbf{T}$ ($\Box\varphi \to \varphi$) to collapse ($\varphi \leftrightarrow \Box\varphi$) depends on classical propositional principles,  so in such systems that implication does not necessarily hold: non-bivalence changes the inevitability of collapse. We will return to this point in our own proposed model below.

Universal bivalence is optional in intuitionistic modal logics, a broad family of modal logics axiomatized uniformly in Liao \cite{liao2023intuitionistic}. In this setting, neither $\mathbf{C}$ nor $\mathbf{T}$ are built in. Specifically, adopting $\mathbf{C}$ alone does not yield the collapse: given the non-bivalent base and the absence of any imposed reflexivity, the equivalence $\varphi\leftrightarrow\Box\varphi$ does not follow. Collapse arises only if one also postulates reflexivity (which is in fact axiom $\mathbf{T}$). Here too a non-bivalent base helps avoid the collapse. 

Collapse is avoided in \cite{lewitzka2017amalgam}, where \emph{amalgam} modal logics are introduced that combine an intuitionistic core with a classical modal layer as one system. Here $\mathbf{T}$ holds, while $\mathbf{C}$ is not valid: there exist models with $\varphi$ true at a world but $\Box\varphi$ false. Thus the amalgam avoids modal collapse.

Model transformation logics provide a dynamic approach. In Fervari et al. \cite{fervari2014relation}, relation-changing modal logics extend the basic modal language with dynamic operators that modify the accessibility relation while evaluating a formula (adding, deleting, or swapping links, locally or globally). Since the truth of $\mathbf{C}$ depends on the accessibility pattern, such updates can move a model from validating $\mathbf{C}$ to refuting it (e.g., by adding a link to a counterexample world), or conversely make $\mathbf{C}$ hold (e.g., by deleting all outgoing links from the current world). Thus these logics provide a dynamic handle on the frame conditions that may determine whether $\mathbf{C}$ holds.

Analogously, Girard and Weber \cite{girard2019modal} develop modal logic in a metatheory without contraction and recover standard frame–formula correspondences under this structural restriction. Their result shows that modal proof theory can be carried out with weakened structural rules, and in such settings, adding $\mathbf{C}$ does not by itself yield $\varphi \leftrightarrow \Box \varphi$ unless further principles (notably axiom $\mathbf{T}$) are also postulated.

More recently, some works have explored non-bivalent or paracomplete modal logics in which Axiom \textbf{C} can be considered without forcing collapse. For instance, Drobyshevich and Wansing develop proof systems for \emph{FDE}-based modal logics (four-valued, paraconsistent/paracomplete) \cite{drobyshevich2020}. Their results show that robust modal reasoning can be carried out over a non-bivalent semantics; in such settings it is easy to see that adopting \textbf{C} does not by itself entail $\varphi \leftrightarrow \Box\varphi$ unless further principles, most notably \textbf{T}, are also postulated. On the constructive side, Dalmonte presents a general framework for \emph{intuitionistic non-normal} modal logics \cite{DalmonteGrelloisOlivetti2020}, providing a non-bivalent base in which neither \textbf{C} nor \textbf{T} is built in; one can thus adopt \textbf{C} while withholding \textbf{T}, so that collapse is not automatic. An algebraic approach is discussed by Figallo-Orellano, Pérez-Gaspar and Ramírez-Contreras, who study modal systems over De Morgan structures that validate paraconsistent/paracomplete behaviour \cite{figallo2022}; these semantics again separate actual truth from necessity, allowing one to formulate Axiom \textbf{C} without collapsing $\Box \varphi$ into $\varphi$. From a proof-theoretic perspective, Hodes develops “one-step” modal calculi in both intuitionistic and classical variants \cite{Hodes2021Part1}, showing that modal proof theory can be designed over non-bivalent bases without building in principles that generate the collapse. Pawłowski and Skurt introduce eight-valued non-deterministic semantics for normal modal logics \cite{pawlowski2024}, where truth values record not only truth/falsity but also (non-)necessity and (non-)possibility; this fine-grained setting makes explicit how \textbf{C} can hold while the equivalence $\varphi \leftrightarrow \Box\varphi$ remains blocked unless additional classical constraints are imposed.

\section{Conjectures} \label{sec:conjectures}

A large part of human reasoning begins neither from complete knowledge nor from total ignorance, but from situations in which much is already established, and something still remains open. This happens in ordinary life, in science, in historical interpretation, in legal argument, and in many other situations: available facts may be enough to constrain what can reasonably be said, while still leaving room for incompatible continuations. Conjectures arise in these situations, in which reality is not empty, but not yet fully determined either.

A conjecture is first of all a subjective cognitive state of an actor: it expresses a cognitive commitment internal to a subject or to a community of subjects. In this sense, it belongs to the doxastic sphere, yet this does not make it an arbitrary belief. Generic beliefs are in principle compatible with almost anything: truths, falsehoods, fantasies, confusions, even contradictions. On the other hand, knowledge is a fairly constrained state, only concerning (some of) what is established to be true. Conjectures are constrained differently: they neither float free from facts nor remain restricted to the factual base as currently available.

In our understanding of the term, conjectures are similar to beliefs in that they suppose something to be true. Yet they are more constrained than arbitrary beliefs, which may range over arbitrary possibilities and may internally assume anything as true, including counterfactuals (e.g., “Suppose Napoleon never existed.”) By conjecture, instead, we mean cognitive states that do not contrast with available facts, but extend them with new propositions that, at the moment, are neither true nor false.

This implies that conjectures make sense only under a non-bivalent setting, i.e., where well-formed statements are not restricted to be either true or false. If every relevant proposition were already fully determined, conjectures would lose their function: conjecturing a proposition already determined as true would add nothing, while conjecturing a proposition already determined as false would amount to contradiction, wishful thinking, or at least delusional subjective belief. 

When some propositions are allowed to remain neither true nor false, a conjecture consists in taking one of these open propositions and determining it within the actor’s cognitive stance, and leaving the rest of the factual base in place. 

The general form of conjectural reasoning begins therefore from a shared base of established facts, which a conjecture extends with one or more propositions that the base still leaves open. This extension now supports further reasoning, consequences can be derived, alternatives can be compared, incompatibilities can be exposed with other conjectures, and practical decisions can be taken. Thus the conjecture does not replace, erase or alter the base, but develops it in one admissible direction rather than another.

Consider for instance \textit{future contingents}. Imagine a sporting event that will take place tomorrow. At present, it is neither a fact that team A will win the match, nor that it will lose: the corresponding proposition ``A wins the match'' is open in the current factual base. Yet reasoning does not need to stop because of this: one may reason under the conjecture that A wins, and, assuming the full set of established facts plus the conjecture, wonder what follows, maybe A winning the championship, eliminating a rival, getting in the range of a particular ranking. Similarly, one may reason under the opposite conjecture of A losing the match using exactly the same factual base  and obtain a different line of consequences. 

A similar structure appears in \textit{historical analysis}, as in, e.g., authorship attributions of works of art. Claims about authorship are rarely plain facts, since they often depend on documentary traces, stylistic judgments, provenance chains, testimonies, expert traditions, and interpretive choices. In \cite{drucker2011factaCapta}, this difference is captured through the distinction between \textit{facta} and \textit{capta}. \textit{Facta} are states of affairs taken to hold independently of interpretation, such as the dimensions of a painting or the material support on which it is executed. \textit{Capta} are what scholars apprehend, select, interpret, and report, including documentary attributions and art-historical claims. Some attributions become so stable that they function, for practical purposes, as settled points in the shared base: for instance, the attribution of the \textit{Mona Lisa} to Leonardo da Vinci is historically unchallenged. Others remain open in a much stronger sense: for instance, the attribution of the \textit{Salvator Mundi} to Leonardo remains controversial to this day, and no less than seven different attributions are adopted by different scholars, all of which hail from the same factual base and extend it in different directions. Scholars reasoning under the conjecture that Leonardo is indeed the author derive consequences about valuation, stylistic coherence, chronology, and ultimately incompatibility with rival attributions, that extend the same evidential base differently. In these situations conjectures do not simply expose the possible consequences of establishing one or more open propositions, but allow divergences, disagreements and ultimately uncertainty to emerge and even to be expressed at all. 

Even in the hard sciences, in the presence of \textit{contrasting empirical evidence}, we can find instances of conjectures. Consider a claim such as ``the BRCA1--BARD1 E3 ligase activity is required for tumor suppression.'' This is not a case in which no relevant experiment has been performed. On the contrary, the literature contains serious and methodologically credible lines of evidence pointing in opposite directions. Some experimental programs have suggested that the ligase activity is not essential \cite{Findlay2023BRCA1Dispensable}, while more recent discussion has argued that its role remains central and, overall, highly controversial \cite{Chaudhuri2023BRCA1Controversy}. Under these conditions, the proposition does not enter the shared factual base as either true or false. It remains open, and researchers may reason under the conjecture that the effect is real, or under the opposite conjecture that it is not, while disagreement persists over the claim itself and over whether one experimental design was misleading, whether biological context matters more than expected, or whether the original question was too coarse-grained. 

Indeed we could possibly apply the term conjectures even to all subjective extensions of reality with propositions that will never be considered facts, as in \textit{subjective opinions}. Consider statements such as ``pizza is better than hot dogs,'' or ``Taylor Swift is the greatest musical artist of this century.'' These are not statements that will ever be found true or false in the future, nor statements whose truth or falsity has been lost in the mist of time, yet in the past were (and still, in some conceivable yet unavailable universe, are) either true or false, but statements that are intrinsically neither true nor false from any objective vantage point. The shared factual base will never contain such propositions, yet subjectively, for the actors who hold them, these statements may be more vivid, more immediate, and more true than many claims they accept via testimony, education, or deference to external authority. For these actors they are definitely settled points from which further choices and decisions can proceed. The personal certainty of the judgment is psychologically strong, and their logical role derives from this subjective certainty, creating a local extension of reality limited to those actors. 

These scenarios differ in content, status, and stakes, but exhibit the same basic structure: there is a factual base that is already rich enough to constrain reasoning, within which propositions remain open. An actor then determines some propositions locally and reasoning proceeds relative to the resulting extension. In future contingents the openness concerns what has not yet happened, in historical attribution it concerns competing disagreements of a shared evidential record, and in subjective opinions it concerns statements that will remain globally open while becoming locally determined for the actor. In all cases, conjectures do more than extend a partial reality: they make uncertainty and disagreement logically visible, because they show how the same incomplete base may open into incompatible but internally consistent extensions.

\section{Conjectural Logics}\label{sec:conjecturalLogics}

Adopting a non-bivalent approach to modal logics, in what follows we assume a base logic—either propositional or description-oriented—that permits well-formed formulas to be undetermined at a world. This condition is satisfied whenever valuations are partial functions over atomic formulas.

In this section we introduce the modal structures built over such a non-bivalent base.

\subsection{Non-Bivalent Bases}\label{subsec:nonbivalent}

Let $\mathcal{L}$ be a propositional language generated from a set of atomic formulas $\mathsf{At}$ by the usual Boolean connectives.

\begin{definition}
A \emph{factual base} is a partial valuation
\[
V : \mathsf{At} \rightharpoonup \{\top, \bot\}.
\]
\end{definition}

\begin{definition}
Given a factual base $V$, its associated induced valuation $\widehat{V}$ is the partial valuation on all formulas of $\mathcal{L}$ induced compositionally from $V$ under the usual Boolean connectives.
\end{definition}

That is, $V$ assigns truth values only to a subset of atomic formulas. The domain of $V$, written $\mathrm{dom}(V)$, represents the set of propositions whose truth value is already determined in the current informational state.

We extend classical evaluation to complex formulas in the following sense.

\begin{definition}
A formula $\varphi$ is determined in $V$ if $\widehat{V}(\varphi)$ is defined. Otherwise, $\varphi$ is undetermined in $V$.
\end{definition}

In intuitive terms, a factual base records what is already determined at the atomic level, while its induced valuation $\widehat{V}$ determines which complex formulas already have a settled truth value. Undetermined formulas correspond to propositions whose truth status has not yet been established.

\begin{definition}
A factual base $V$ is total if every atomic formula in $\mathsf{At}$ belongs to $\mathrm{dom}(V)$. It is non-total otherwise.
\end{definition}

If $V$ is total, every formula in $\mathcal{L}$ receives a classical truth value. If conjectures are understood as extensions of a factual base that preserve what is already determined, then the base must be non-total. If the base were total, every atomic formula would already have a truth value, and any extension preserving all determined truths would coincide with the base itself. Hence conjectural extensions would be vacuous.

\bigskip

Although factual bases are partial, their evaluation is classical wherever defined.

Factual bases can be compared in terms of informational extension: enlarging the domain of determination while preserving previously established values induces a natural ordering among bases. Indeed, the family of factual bases admits a natural partial order defined by inclusion:
\[
V_1 \subseteq V_2
\quad \text{iff} \quad
\mathrm{dom}(V_1) \subseteq \mathrm{dom}(V_2)
\text{ and }
V_2(p) = V_1(p)
\text{ for all } p \in \mathrm{dom}(V_1).
\]

\subsection{Kripke Structures with a Non-Bivalent Base}\label{subsec:accessibility}

A modal logic extends a base logic by introducing a unary modal operator $\Box$.
Its semantics is defined in terms of a Kripke frame.
In the present setting, the underlying propositional semantics is non-bivalent, i.e.,
valuations at worlds are partial functions $V_w : \mathsf{At} \rightharpoonup \{\top,\bot\}$.

What we discussed previously allows us to assume that:
\begin{itemize}
\item Each world $w$ is equipped with a factual base $V_w : \mathsf{At} \rightharpoonup \{\top,\bot\}$ in the sense of Section~\ref{subsec:nonbivalent}.
\item Satisfaction is defined only for formulas whose value is defined by the induced valuation $\widehat{V}_w$, and evaluation is classical wherever defined.
\end{itemize}

\begin{definition}
A \textbf{Kripke frame} is a pair $\langle W, R \rangle$, where:
\begin{itemize}
\item $W$ is a non-empty set of worlds.
\item $R \subseteq W \times W$ is a binary accessibility relation.
\end{itemize}
\end{definition}

\begin{definition}
A \textbf{Kripke model} is a triple
$\mathcal{M} = \langle W, R, \mathcal{V} \rangle$,
where $\langle W,R\rangle$ is a frame and $\mathcal{V}$ assigns to each world $w \in W$ a partial valuation $V_w : \mathsf{At} \rightharpoonup \{\top,\bot\}$.
\end{definition}

Thus, for each world $w \in W$, we write $V_w = \mathcal{V}(w)$. Its associated induced valuation on all formulas is denoted by $\widehat{V}_w$.
Satisfaction is restricted to determined truth:
\begin{equation}
\mathcal{M},w \models \varphi
\quad \text{iff} \quad
\varphi \text{ is determined at } w
\text{ and }
\widehat{V}_w(\varphi)=\top.
\end{equation}

Only truth is designated, and if a formula is undetermined at $w$, then neither it nor its negation is satisfied at $w$.
The modal operator is interpreted in the standard way:
\begin{equation}
\mathcal{M}, w \models \Box \varphi
\quad \text{iff} \quad
\forall w' \in W,\
\bigl(w R w' \Rightarrow \mathcal{M}, w' \models \varphi \bigr).
\end{equation}

\paragraph{Modal Axioms.}

We recall the usual modal schemata and their associated frame conditions:
\begin{itemize}
\item \textbf{(K)} $\Box(\varphi \rightarrow \psi) \rightarrow (\Box \varphi \rightarrow \Box \psi)$, valid in all Kripke frames.
\item \textbf{(D)} $\Box \varphi \rightarrow \neg \Box \neg \varphi$, corresponding to seriality.
\item \textbf{(T)} $\Box \varphi \rightarrow \varphi$, corresponding to reflexivity.
\item \textbf{(4)} $\Box \varphi \rightarrow \Box \Box \varphi$, corresponding to transitivity.
\item \textbf{(5)} $\neg \Box \varphi \rightarrow \Box \neg \Box \varphi$, corresponding to Euclideanness.
\end{itemize}

These axioms correspond to well-known frame properties in the standard classification of modal logics \cite{chagrov1997modal, blackburn2001modal}.

\paragraph{Validity.}

Let $\mathcal{M}=\langle W,R,\mathcal{V}\rangle$ be a Kripke model.
\begin{itemize}
\item $\mathcal{M},w\models\varphi$ if $\varphi$ is determined and true at $w$.
\item $\varphi$ is valid in $\mathcal{M}$ if $\mathcal{M},w\models\varphi$ for all $w\in W$.
\item $\varphi$ is valid on a frame $\langle W,R\rangle$ if it is valid in every model built on that frame.
\end{itemize}

\paragraph{Provability.}

Given a deductive system $\Sigma$, we write $\Sigma\vdash\varphi$ to mean that $\varphi$ is provable in $\Sigma$, and $\Gamma\vdash_\Sigma\varphi$ if $\varphi$ is provable from $\Gamma$ in $\Sigma$.
Uniform Substitution and the Deduction Theorem are assumed.

\bigskip

In a modal logic based on a non-bivalent semantics, not all formulas receive a definite truth value at every world. This necessitates a distinction between formulas that are \emph{defined} at a world and those that are not.

The following notions make explicit, at the level of worlds in a Kripke model, the notions of determined formula and informational extension introduced for factual bases in Section~\ref{subsec:nonbivalent}.

\begin{definition}[Defined Formulas]
Let $\mathcal{M}$ be a Kripke model and $w$ a world in $\mathcal{M}$. The set of \textit{defined formulas} in $w$ is:
\[
\mathcal{D}^\mathcal{M}_w
=
\left\{\, \varphi\ \middle|\ 
\widehat V_w(\varphi)\ \text{is defined}
\,\right\}.
\]
This is equivalent to $\varphi \in \mathcal{D}^\mathcal{M}_w$ iff $\mathcal{M}, w \models \varphi$ or $\mathcal{M}, w \models \neg \varphi$.
\end{definition}

Thus a formula is defined at $w$ precisely when it is determined there, that is, when $\widehat{V}_w(\varphi)$ is defined, and equivalently when either it or its negation is satisfied at $w$. The factual base $V_w$ determines truth values for atomic formulas, and the associated induced valuation $\widehat{V}_w$ extends this information compositionally to complex formulas wherever defined.
When the model $\mathcal{M}$ is clear from context, we omit it and simply write $\mathcal{D}_w$.

\begin{definition}[Definedness-Preserving Extension]\label{def:definedness-preserving-extension}
A world $w'$ is a definedness-preserving extension of $w$ if, in the inclusion order on factual bases introduced in Section~\ref{subsec:nonbivalent},
\[
V_w \subseteq V_{w'}.
\]
\end{definition}

\begin{definition}[Definedness-Preserving Accessibility]\label{def:definedness-preserving-accessibility}
An accessibility relation $R$ preserves definedness if every accessible world is a definedness-preserving extension of its predecessor, that is, if
\[
w R w' \quad \Rightarrow \quad V_w \subseteq V_{w'}.
\]
\end{definition}

This constraint ensures that accessible worlds may assign truth values to additional formulas but cannot contradict already determined ones. Thus, when accessibility preserves definedness, the modal operator $\Box$ guarantees stability of determined truth under accessibility. A formula $\Box \varphi$ holds at $w$ iff $\varphi$ holds in all worlds $w'$ such that $w R w'$.

\subsection{Axiom \textbf{C}}\label{subsec:axiomC}

We define Axiom \textbf{C} as the following modal schema:
\begin{equation}
\textbf{(C)} \quad \varphi \rightarrow \Box \varphi
\end{equation}

As discussed previously, in classical modal logic this axiom is rarely adopted, since it leads directly to modal collapse when combined with Axiom \textbf{T} ($\Box \varphi \rightarrow \varphi$). However, in our setting, without \textbf{T} and based on a non-bivalent base logic, the collapse is not guaranteed, and the behavior of Axiom \textbf{C} must be reassessed accordingly.

In Kripke-style semantics, $\Box \varphi$ is interpreted as: ``$\varphi$ holds in all accessible worlds.'' Formally, for a model $\mathcal{M} = \langle W, R, \mathcal{V} \rangle$ and a world $w \in W$, this means:
\begin{equation}
\mathcal{M}, w \models \Box \varphi \quad \text{iff} \quad \forall w' \text{ such that } w R w',\ \mathcal{M}, w' \models \varphi
\end{equation}

Under \textbf{C}, by the semantics of implication, $\varphi \rightarrow \Box \varphi$ is satisfied at a world $w$ if \textit{either} $\varphi$ is not satisfied at $w$, \textit{or} $\Box \varphi$ is satisfied at $w$. In other words, satisfaction of $\varphi \rightarrow \Box \varphi$ at $w$ guarantees that whenever $\varphi$ is satisfied at $w$, then $\Box \varphi$ is also satisfied at $w$:
\begin{equation}
\text{If } \mathcal{M}, w \models \varphi, \text{ then } \mathcal{M}, w \models \Box \varphi
\end{equation}

Now, applying the semantics of the $\Box$ operator, this expands to the following condition:
\begin{equation}
\text{If } \mathcal{M}, w \models \varphi, \text{ then } \forall w' \text{ such that } w R w',\ \mathcal{M}, w' \models \varphi
\end{equation}

That is, if $\mathcal{M}, w \models \varphi$, then for every $w'$ such that $w R w'$, it must also hold that $\mathcal{M}, w' \models \varphi$. In our framework, this propagation is not enforced by properties of $R$ alone, such as reflexivity or transitivity, but by the interaction between $R$ and the valuation structure of the model. More precisely, what Axiom \textbf{C} demands is that any formula satisfied at $w$ remain satisfied along all accessible worlds. 

This motivates the following:

\begin{definition}
A Kripke model $\mathcal{M}$ satisfies the \textbf{C-propagation condition} at $w$ if for every formula $\varphi$ such that $\mathcal{M}, w \models \varphi$, it holds that $\mathcal{M}, w' \models \varphi$ for all $w'$ such that $w R w'$.
\end{definition}

This condition amounts to a form of truth preservation: formulas that evaluate to $\top$ in $w$ must remain $\top$ in all accessible worlds. Since satisfaction is defined relative to the designated value $\top$, and undefinedness is not designated, Axiom \textbf{C} never forces previously undetermined formulas to become determined. It requires only preservation of already satisfied formulas.

Thus Axiom \textbf{C} governs structural coherence between a world and its accessible extensions: no accessible state can retract truths already evaluated as $\top$. This interpretation is compatible with non-bivalent semantics, since it never requires global determination, but only stability of designated truth. The formula $\varphi \rightarrow \Box \varphi$ therefore expresses a preservation constraint: whenever $\varphi$ holds at a world, it must continue to hold throughout its accessible structure.

This role becomes especially meaningful in non-bivalent logics: since truth and falsity are not exhaustive, the inference from $\varphi$ to $\Box \varphi$ stipulates propagation of designated truth where it is already present. As such, the inclusion of \textbf{C} does not by itself trigger collapse. Collapse arises only if \textbf{C} is combined with \textbf{T}, as discussed below.

\begin{definition}
We call $\mathcal{C}$ the class of Kripke models such that accessibility preserves definedness and satisfies C-propagation.
\end{definition}

\textbf{N.B.}: The condition defining $\mathcal{C}$ is a valuation-propagation constraint, not a frame property. Since satisfaction is defined by $\mathcal{M},w\models\varphi$ iff $\widehat V_w(\varphi)=\top$, this formulation is equivalent to the semantic reading of Axiom \textbf{C} adopted in the chapter.

\begin{proposition}\label{prop:C-implies-4}
\[
\textbf{C} \ \vdash \textbf{4} \text{, and } \textbf{C} \ \vdash \textbf{5}
\]
\end{proposition}

\begin{proof}
Since \textbf{C} ($\varphi \rightarrow \Box \varphi$) applies to every formula, instantiate it with $\varphi := \Box \psi$. This immediately yields $\Box \psi \rightarrow \Box \Box \psi$, which is the formula usually called Axiom \textbf{4}.

Since \textbf{C} ($\varphi \rightarrow \Box \varphi$) applies to every formula, instantiate it with $\varphi := \neg \Box \psi$. This immediately yields $\neg \Box \psi \rightarrow \Box \neg \Box \psi$, which is the formula usually called Axiom \textbf{5}.
\end{proof}

Thus, under Axiom \textbf{C}, axioms \textbf{4} and \textbf{5} are not independent principles, but are direct applications of it.

\subsection{Interaction of Axiom \textbf{C} with Standard Modal Axioms}\label{subsec:interactionC}

Axiom \textbf{C} can be consistently added to a variety of modal systems. However, its interaction with other axioms, especially \textbf{T}, can introduce semantic consequences that range from non-trivial coherence constraints to full modal collapse. In this section, we analyze these interactions with respect to frame conditions and model behavior, under the assumption that the base logic is non-bivalent.

\paragraph{C with K.}
The minimal modal logic \textbf{K} requires only that $\Box$ distributes over implication (Axiom \textbf{K}). Since \textbf{C} imposes no specific structural constraints on the accessibility relation $R$, it is consistent with \textbf{K}. In this setting, $\Box \varphi$ denotes propagation of satisfaction across accessible worlds, and Axiom \textbf{C} ensures that satisfaction of $\varphi$ at any world $w$ must propagate through $R$. The resulting logic, \textbf{KC}, is characterized semantically by closure under $R$: if $\mathcal{M}, w \models \varphi$, then $\mathcal{M}, w' \models \varphi$ for all $w'$ such that $w R w'$. This is not a frame condition in the usual sense, but a valuation propagation constraint. The system \textbf{KC} is thus well-formed and non-trivial.

\paragraph{C with D.}
Adding axiom \textbf{D}, $\Box \varphi \rightarrow \neg \Box \neg \varphi$, requires that the accessibility relation $R$ be serial: for every world $w$, there exists at least one $w'$ such that $w R w'$. This guarantees that $\Box \varphi$ cannot be satisfied vacuously. When added to \textbf{KC}, i.e.\ \textbf{KDC}, Axiom \textbf{C} ensures propagation of designated truth, while Axiom \textbf{D} prevents trivial modal validity. In a non-bivalent setting, this combination ensures that $\Box \varphi$ is evaluated over a non-empty image of $R$, and that propagation under \textbf{C} is never vacuous.

\paragraph{C with T: modal collapse.}
Axiom \textbf{T}, $\Box \varphi \rightarrow \varphi$, enforces reflexivity on $R$: each world must access itself. When added to \textbf{C}, it yields the bi-implication $\varphi \leftrightarrow \Box \varphi$, thereby erasing the modal distinction between local satisfaction and modal propagation. In classical modal logic, this configuration leads to modal collapse. The same structural collapse occurs in the present setting: if both \textbf{C} and \textbf{T} hold, necessity coincides with local satisfaction. For this reason, we explicitly avoid adopting \textbf{T}.

\paragraph{C with 4 and 5.}
Axiom \textbf{4}, $\Box \varphi \rightarrow \Box \Box \varphi$, corresponds in the standard setting to transitivity on $R$. Axiom \textbf{5}, $\neg \Box \varphi \rightarrow \Box \neg \Box \varphi$, corresponds similarly to Euclideanness. In the present framework, however, these principles no longer function as independent enrichments once \textbf{C} is assumed. As shown in Section~\ref{subsec:axiomC}, both are immediate consequences of \textbf{C} by uniform substitution. They may therefore still be read as clarifications of the propagation pattern induced by \textbf{C}, but they do not generate distinct systems beyond those already determined by the presence of \textbf{C} itself.

\subsection{The Logics KC and KDC}\label{subsec:logicsKCetc}

Building on the semantics of Axiom \textbf{C}, we define two modal systems by combining \textbf{C} with standard axioms that constrain the accessibility relation. These logics extend normal modal logic over a non-bivalent base, while avoiding reflexivity, which leads to collapse.

The system \textbf{KC} consists of the modal logic \textbf{K}, which includes the distribution axiom and necessitation, together with Axiom \textbf{C}. It characterizes models where designated truth propagates across accessible worlds. Since no structural condition is imposed on $R$, \textbf{KC} is compatible with arbitrary accessibility, provided that C-propagation is satisfied.

\textbf{KDC} adds Axiom \textbf{D}, ensuring seriality. This guarantees that every world has at least one accessible successor, preventing vacuous satisfaction of modal formulas. Together with Axiom \textbf{C}, this ensures that propagation of designated truth always occurs over a non-empty domain.

Since Axiom \textbf{C} immediately yields both Axioms \textbf{4} and \textbf{5}, there are no additional systems \textbf{KC45} or \textbf{KDC45} over and above \textbf{KC} and \textbf{KDC}. Any reference to \textbf{KC45} or \textbf{KDC45} is therefore only a notational reminder that the corresponding instances of \textbf{4} and \textbf{5} are already available once \textbf{C} is active.

These systems define a family of modal logics governed by non-bivalent evaluation and propagation of designated truth. They provide a structured framework for reasoning in settings where local satisfaction does not exhaust modal validity.

\subsection{Soundness and completeness}\label{subsec:soundness}

We assume a propositional base $B$, sound and complete, admitting Uniform Substitution and the Deduction Theorem. Valuations in $B$ are partial and consistent, and whenever formulas are determined, their semantic evaluation and proof-theoretic behavior coincide with ordinary classical propositional reasoning.

A Kripke model is a triple $\mathcal{M}=\langle W,R,\mathcal{V}\rangle$, where $W$ is a nonempty set of worlds, $R\subseteq W\times W$ is the accessibility relation, and $\mathcal{V}$ assigns to each world $w$ a partial valuation $V_w$ on atoms. Each $V_w$ induces compositionally a partial valuation $\widehat V_w$ on formulas. We write $\mathcal{M},w\models\varphi$ for truth at $w$.

\subsubsection{Soundness and completeness of \textbf{KC}}

\begin{definition}
A $\mathbf{KC}$-\emph{theory} is a set of formulas closed under derivability in $\mathbf{KC}$. A $\mathbf{KC}$-theory is \emph{consistent} if there is no formula $\varphi$ such that both $T \vdash_{\mathbf{KC}} \varphi$ and $T \vdash_{\mathbf{KC}} \neg \varphi$. A maximal $\mathbf{KC}$-consistent theory is a $\mathbf{KC}$-consistent theory maximal under inclusion.
\end{definition}

\begin{definition}
A set of formulas $s$ is $\mathbf{KC}$-consistent if there is no formula $\varphi$ such that both $s \vdash_{\mathbf{KC}} \varphi$ and $s \vdash_{\mathbf{KC}} \neg\varphi$.
\end{definition}

\begin{definition}[Canonical model for \textbf{KC}]
Let $\mathbf{KC}$ be an axiomatic system with base $B$, Axiom $\mathbf{K}$, the Rule of Necessitation, Axiom $\mathbf{C}$, and Uniform Substitution. We define the canonical model $\mathfrak{M}^{\mathrm{can}} = (W^{\mathrm{can}}, R^{\mathrm{can}}, \mathcal{V})$ as follows:
\begin{itemize}
\item $W^{\mathrm{can}}$ is the set of maximal $\mathbf{KC}$-consistent theories.
\item $R^{\mathrm{can}}$ is the accessibility relation such that $w \, R^{\mathrm{can}} \, w'$ iff for every $\Box\varphi\in w$ we have $\varphi\in w'$.
\item For atoms, $V_w(p)=\top$ iff $p\in w$, $V_w(p)=\bot$ iff $\neg p\in w$, and $V_w(p)$ is undefined otherwise.
\end{itemize}
\end{definition}

For each world $w$, the induced valuation $\widehat V_w$ is obtained compositionally from $V_w$ according to the non-modal clauses of $B$, and the modal operator $\Box$ is interpreted in the standard Kripke way.

\begin{lemma}\label{lem:lindenbaum}
Every $\mathbf{KC}$-consistent set $s$ extends to some $w\in W^{\mathrm{can}}$ with $s\subseteq w$, where $w$ is a maximal $\mathbf{KC}$-consistent theory.
\end{lemma}

\begin{proof}
Let $\mathcal{S}$ be the family of $\mathbf{KC}$-consistent theories extending the $\mathbf{KC}$-theory generated by $s$, ordered by inclusion. This family is nonempty, since the theory generated by $s$ is consistent whenever $s$ is $\mathbf{KC}$-consistent. Let $\mathcal{S}^*$ be a chain in $\mathcal{S}$, and let
\[
U=\bigcup_{T\in\mathcal{S}^*} T.
\]
Then $U$ is again a $\mathbf{KC}$-theory: if $\Gamma\subseteq U$ is finite and $\Gamma\vdash_{\mathbf{KC}}\theta$, then all formulas in $\Gamma$ already belong to some $T\in\mathcal{S}^*$, since $\mathcal{S}^*$ is totally ordered; because $T$ is a theory, $\theta\in T\subseteq U$. Moreover, $U$ is consistent, since any inconsistency would already appear in some member of the chain. Thus $U$ is an upper bound of $\mathcal{S}^*$. By Zorn’s Lemma, $\mathcal{S}$ has a maximal element $w$, which is a maximal $\mathbf{KC}$-consistent theory extending $s$.
\end{proof}

\begin{lemma}[Modal lifting]\label{lem:modal-lifting}
If $\chi_1,\dots,\chi_n \vdash_{\mathbf{KC}} \psi$, then
\[
\Box\chi_1,\dots,\Box\chi_n \vdash_{\mathbf{KC}} \Box\psi.
\]
\end{lemma}

\begin{proof}
By repeated applications of the Deduction Theorem,
\[
\vdash_{\mathbf{KC}} \chi_1 \to (\chi_2 \to \cdots (\chi_n \to \psi)\cdots).
\]
By Necessitation,
\[
\vdash_{\mathbf{KC}} \Box(\chi_1 \to (\chi_2 \to \cdots (\chi_n \to \psi)\cdots)).
\]
By repeated applications of Axiom $\mathbf{K}$, we derive
\[
\vdash_{\mathbf{KC}} \Box\chi_1 \to (\Box\chi_2 \to \cdots (\Box\chi_n \to \Box\psi)\cdots).
\]
The conclusion follows by repeated Modus Ponens.
\end{proof}

\begin{lemma}\label{lem:successor}
If $\neg\Box\psi \in w$, then there exists $w'$ such that $w R^{\mathrm{can}} w'$ and $\neg\psi \in w'$.
\end{lemma}

\begin{proof}
Define
\[
s=\{\chi \mid \Box\chi \in w\}\cup\{\neg\psi\}.
\]
Assume for contradiction that $s$ is inconsistent. Then there exist finitely many $\chi_1,\dots,\chi_n$ such that
\[
\chi_1,\dots,\chi_n,\neg\psi \vdash_{\mathbf{KC}} \bot.
\]
By the Deduction Theorem and the classical proof-theoretic behavior of $B$ on determined formulas,
\[
\chi_1,\dots,\chi_n \vdash_{\mathbf{KC}} \psi.
\]
By Modal lifting,
\[
\Box\chi_1,\dots,\Box\chi_n \vdash_{\mathbf{KC}} \Box\psi.
\]
Since all $\Box\chi_i$ belong to $w$ and $w$ is a theory, we conclude $\Box\psi \in w$, contradicting $\neg\Box\psi\in w$. Hence $s$ is consistent. By Lemma~\ref{lem:lindenbaum}, $s$ extends to a maximal $\mathbf{KC}$-consistent theory $w'$. By construction, $w R^{\mathrm{can}} w'$ and $\neg\psi \in w'$.
\end{proof}

\begin{lemma}[Truth Lemma]\label{lem:truth}
For every world $w\in W^{\mathrm{can}}$ and every formula $\varphi$:
\begin{enumerate}
\item if $\varphi\in w$, then $\mathfrak{M}^{\mathrm{can}}, w \models \varphi$;
\item if $\neg\varphi\in w$, then $\mathfrak{M}^{\mathrm{can}}, w \models \neg\varphi$;
\item if $\mathfrak{M}^{\mathrm{can}}, w \models \varphi$, then $\varphi\in w$;
\item if $\mathfrak{M}^{\mathrm{can}}, w \models \neg\varphi$, then $\neg\varphi\in w$.
\end{enumerate}
\end{lemma}

\begin{proof}
Clauses (1) and (2) are proved by induction on the structure of $\varphi$.

The atomic case follows immediately from the definition of $V_w$. The Boolean cases follow from the induction hypothesis and the compositional clauses of $B$.

For the modal positive case in clause (1), suppose $\Box\varphi\in w$. Let $w'$ be any world such that $w R^{\mathrm{can}} w'$. By definition of $R^{\mathrm{can}}$, $\varphi\in w'$. By induction hypothesis, $\mathfrak{M}^{\mathrm{can}}, w' \models \varphi$. Hence $\mathfrak{M}^{\mathrm{can}}, w \models \Box\varphi$.

For the modal negative case in clause (2), suppose $\neg\Box\varphi\in w$. By Lemma~\ref{lem:successor}, there exists $w'$ such that $w R^{\mathrm{can}} w'$ and $\neg\varphi\in w'$. By induction hypothesis, $\mathfrak{M}^{\mathrm{can}}, w' \models \neg\varphi$, hence $\mathfrak{M}^{\mathrm{can}}, w' \not\models \varphi$. Therefore $\mathfrak{M}^{\mathrm{can}}, w \models \neg\Box\varphi$.

We now prove clauses (3) and (4) by contradiction using maximality and closure under derivability.

For clause (3), assume $\mathfrak{M}^{\mathrm{can}}, w \models \varphi$ but $\varphi \notin w$. Since $w$ is a maximal consistent theory, the theory generated by $w \cup \{\varphi\}$ cannot be consistent. Hence there exist a finite $\Gamma \subseteq w$ and a formula $\psi$ such that
\[
\Gamma,\varphi \vdash_{\mathbf{KC}} \psi
\qquad\text{and}\qquad
\Gamma,\varphi \vdash_{\mathbf{KC}} \neg\psi.
\]
By the Deduction Theorem,
\[
\Gamma \vdash_{\mathbf{KC}} \varphi\to\psi
\qquad\text{and}\qquad
\Gamma \vdash_{\mathbf{KC}} \varphi\to\neg\psi.
\]
Because $w$ is a theory and $\Gamma\subseteq w$, it follows that
\[
\varphi\to\psi \in w
\qquad\text{and}\qquad
\varphi\to\neg\psi \in w.
\]
By clauses (1) and (2),
\[
\mathfrak{M}^{\mathrm{can}},w\models\varphi\to\psi
\qquad\text{and}\qquad
\mathfrak{M}^{\mathrm{can}},w\models\varphi\to\neg\psi.
\]
Together with $\mathfrak{M}^{\mathrm{can}},w\models\varphi$, we conclude
\[
\mathfrak{M}^{\mathrm{can}},w\models\psi
\qquad\text{and}\qquad
\mathfrak{M}^{\mathrm{can}},w\models\neg\psi,
\]
a contradiction. Therefore $\varphi\in w$.

Clause (4) follows by applying clause (3) to the formula $\neg\varphi$.
\end{proof}

\begin{lemma}\label{lem:M^canInC}
The canonical model $\mathfrak{M}^{\mathrm{can}}$ belongs to $\mathcal{C}$.
\end{lemma}

\begin{proof}
Let $w R^{\mathrm{can}} w'$.

For the first condition of $\mathcal C$, suppose $\widehat V_w(\varphi)=\top$. By the semantic convention of the chapter, this means $\mathfrak M^{\mathrm{can}},w \models \varphi$. By clause (3) of the Truth Lemma~\ref{lem:truth}, $\varphi \in w$. Since $\varphi\to\Box\varphi$ is an axiom of $\mathbf{KC}$ and $w$ is a theory, conclude $\Box\varphi \in w$. By the definition of $R^{\mathrm{can}}$, conclude $\varphi \in w'$. By clause (1) of the Truth Lemma, conclude $\mathfrak M^{\mathrm{can}},w' \models \varphi$. Hence $\widehat V_{w'}(\varphi)=\top$.

For the second condition, if $V_w(p)=\top$, then $p\in w$, hence by $\mathbf C$, $\Box p\in w$, hence $p\in w'$, so $V_{w'}(p)=\top$. If $V_w(p)=\bot$, then $\neg p\in w$, hence by $\mathbf C$ instantiated with $\neg p$, $\Box\neg p\in w$, hence $\neg p\in w'$, so $V_{w'}(p)=\bot$.
\end{proof}

\begin{theorem}[Soundness]
Every theorem of $\mathbf{KC}$ is valid on every model in $\mathcal{C}$.
\end{theorem}

\begin{proof}
The axioms of the base logic $B$ are valid by assumption. Axiom $\mathbf{K}$ and Necessitation are sound under standard Kripke semantics. For Axiom $\mathbf{C}$, if $\mathcal M,w \models \varphi$, then by the definition of $\mathcal C$ every accessible world preserves $\widehat V_w(\varphi)=\top$, hence $\mathcal M,w \models \Box\varphi$. Therefore $\varphi\to\Box\varphi$ is valid on every model in $\mathcal C$. Closure under Modus Ponens yields soundness.
\end{proof}

\begin{theorem}[Completeness]
If $s$ is $\mathbf{KC}$-consistent, then there exist $\mathcal{M}\in\mathcal{C}$ and $w$ such that $\mathcal{M},w\models s$.
\end{theorem}

\begin{proof}
By Lemma~\ref{lem:lindenbaum}, extend $s$ to a maximal $\mathbf{KC}$-consistent theory $w$. By the Truth Lemma~\ref{lem:truth}, every formula of $s$ is satisfied at $w$ in the canonical model. By Lemma~\ref{lem:M^canInC}, the canonical model belongs to $\mathcal C$.
\end{proof}

\subsubsection{Extension to $\mathbf{KDC}$}

The canonical model for $\mathbf{KDC}$ is defined analogously, using maximal $\mathbf{KDC}$-consistent theories. To obtain the $\mathbf{KDC}$ version it remains to prove seriality.

\begin{lemma}
Let $w$ be a maximal $\mathbf{KDC}$-consistent theory, and let
\[
s=\{\chi \mid \Box\chi \in w\}.
\]
Then there exists $w'$ such that $w R^{\mathrm{can}} w'$.
\end{lemma}

\begin{proof}
Assume $s$ inconsistent. Then for some finite $\chi_1,\dots,\chi_n$, derive
\[
\chi_1,\dots,\chi_n \vdash_{\mathbf{KDC}} \bot.
\]
By Modal lifting derive
\[
\Box\chi_1,\dots,\Box\chi_n \vdash_{\mathbf{KDC}} \Box\bot.
\]
Since all $\Box\chi_i$ are in $w$ and $w$ is a theory, conclude $\Box\bot \in w$. By $\mathbf D$ instantiated with $\bot$,
\[
\Box\bot \to \neg\Box\neg\bot.
\]
Since $w$ is a theory, it follows that $\neg\Box\neg\bot \in w$. On the other hand, $\neg\bot$ is provable in the base logic, hence by Necessitation conclude that $\Box\neg\bot$ is a theorem of $\mathbf{KDC}$. Therefore $\Box\neg\bot \in w$. This contradicts consistency. Therefore $s$ is consistent. By the corresponding Lindenbaum construction for $\mathbf{KDC}$-consistent sets, extend $s$ to $w'$. Then conclude $w R^{\mathrm{can}} w'$.
\end{proof}

Soundness for $\mathbf{KDC}$ is obtained by combining the soundness argument for $\mathbf{KC}$ with seriality. Completeness for $\mathbf{KDC}$ is obtained by the same canonical construction, now supplemented by the seriality lemma above. Hence the canonical construction yields soundness and completeness for $\mathbf{KDC}$ over the serial models in $\mathcal{C}$.

\subsection{Modal Collapse under Non-Bivalence}

In classical modal settings Axiom \textbf{C} leads to modal collapse. The purpose of this section is to determine the minimal structural conditions under which \textbf{C} does \emph{not} lead to collapse.
We find that the decisive factor, besides the absence of Axiom \textbf{T}, is that a non-bivalent base is chosen and that non-bivalence is concretely realized in the model.

\begin{definition}
We say that a Kripke model $\mathcal M$ is admissible if it has a non-bivalent partial valuation, truth as the only designated value, serial accessibility, compositional propagation of undeterminedness
(i.e.\ if a subformula is undetermined at a world, then every compound formula containing it is undetermined at that world), and no built-in validity of Axiom \textbf{T}.
\end{definition}

\begin{definition}
We say that a Kripke model $\mathcal{M}$ is concretely non-bivalent if there is a world $w$ in $\mathcal{M}$ that concretely contains at least one formula $\varphi$ that is undetermined, i.e., 
\[
\exists w\in W\,\exists\varphi\ \bigl(\mathcal M,w\not\models\varphi\ \wedge\ \mathcal M,w\not\models\neg\varphi\bigr).
\]
\end{definition}

\begin{definition}
We say that a Kripke model $\mathcal{M}$ is not-collapsing if
\[
\exists w\in W\,\exists\varphi\ \bigl(\mathcal M,w\not\models(\varphi\leftrightarrow \Box\varphi)\bigr).
\]
\end{definition}

\begin{theorem}\label{thm:C-collapse-iff-concrete-nonbivalence}
In every admissible $\mathcal M$, if $\mathcal M\models \mathbf C$ then
\[
\mathcal M \text{ is not-collapsing}
\quad\text{iff}\quad
\mathcal M \text{ is concretely non-bivalent}.
\]
\end{theorem}

\begin{proof}
Assume $\mathcal M$ is admissible and $\mathcal M\models \mathbf C$.

\smallskip
\noindent\textbf{($\Leftarrow$)} Suppose $\mathcal M$ is concretely non-bivalent. Then there exist $w$ and $\varphi$ such that
\[
\mathcal M,w\not\models\varphi
\quad\text{and}\quad
\mathcal M,w\not\models\neg\varphi.
\]
Hence $\varphi$ is undetermined at $w$. By propagation of undeterminedness, the formula $(\varphi\leftrightarrow \Box\varphi)$ is undetermined at $w$ as well, and therefore it is not satisfied at $w$. Thus $\mathcal M$ is not-collapsing.

\smallskip
\noindent\textbf{($\Rightarrow$)} Suppose $\mathcal M$ is not-collapsing. Then there exist $w$ and $\varphi$ such that
\[
\mathcal M,w\not\models(\varphi\leftrightarrow \Box\varphi).
\]
We show that $\varphi$ must be undetermined at $w$, i.e.\ neither $\varphi$ nor $\neg \varphi$ is satisfied there. Assume towards a contradiction that $\varphi$ is determined at $w$.

\smallskip
\noindent
If $\mathcal M,w\models\varphi$, then since $\mathcal M\models \mathbf C$ we have $\mathcal M,w\models(\varphi\to \Box\varphi)$, hence $\mathcal M,w\models \Box\varphi$. Therefore both directions $(\varphi\to \Box\varphi)$ and $(\Box\varphi\to \varphi)$ are satisfied at $w$, and so $\mathcal M,w\models(\varphi\leftrightarrow \Box\varphi)$, contradiction.

\smallskip
\noindent
If instead $\mathcal M,w\not\models\varphi$ and $\varphi$ is determined at $w$, then by classical evaluation where defined we have $\mathcal M,w\models\neg\varphi$. By $\mathbf C$ instantiated with $\neg\varphi$, we get $\mathcal M,w\models \Box\neg\varphi$. Since accessibility is serial, pick some $w'$ with $wRw'$. Then $\mathcal M,w'\models\neg\varphi$, hence $\mathcal M,w'\not\models\varphi$, so $\mathcal M,w\not\models\Box\varphi$. Now $\varphi$ is false at $w$ and $\Box\varphi$ is false at $w$, so both implications $(\varphi\to \Box\varphi)$ and $(\Box\varphi\to \varphi)$ are satisfied at $w$, hence again $\mathcal M,w\models(\varphi\leftrightarrow \Box\varphi)$, contradiction.

\smallskip
\noindent
Thus $\varphi$ cannot be determined at $w$. Therefore
\[
\mathcal M,w\not\models\varphi
\quad\text{and}\quad
\mathcal M,w\not\models\neg\varphi,
\]
so $\mathcal M$ is concretely non-bivalent.
\end{proof}

The theorem shows that, in a serial system validating \textbf{C}, avoidance of collapse depends on the existence of at least one world and one formula for which neither the formula nor its negation is satisfied. If satisfaction is effectively bivalent at every world, then no such counterexample remains available, and the model collapses to the equivalence $\varphi \leftrightarrow \Box\varphi$. Effective non-bivalence is therefore the minimal structural condition that allows \textbf{C} to be adopted without modal collapse.

\section{The inclusion theorem of cognitive logics}\label{sec:inclusiontheorem}

Section~\ref{sec:conjecturalLogics} established conjectural logics as modal systems based on a non-bivalent Kripke semantics, and proved their main formal properties. The current section situates them within a broader set-theoretic picture that also includes the standard doxastic and epistemic families.

Cognitive modal logics admit a common reading where worlds are treated as valuation sets ordered by inclusion. Accordingly, the key structure is the relation between a designated reality world, a shared background of accepted information, and the cognitive worlds associated with different modal attitudes. This makes it possible to describe how doxastic, epistemic, and conjectural logics are related and where they diverge.

This section introduces that set-theoretic representation and studies the constraints corresponding to the axioms D, T, and C. The main result is the \textit{inclusion theorem}, which provides a general set-theoretic characterization of conjectural, doxastic, and epistemic systems. 

\subsection{A set-theoretical characterization of cognitive logics}
In the following, let $\mathbb{P}$ be a non-empty set of atoms. 

\begin{definition}[Cognitive world]\label{def:world}
A cognitive world is a set $w\subseteq \mathbb{P} \times\{\top,\bot\}$. We write $w\models p$ if $(p,\top)\in w$ and $w\models\neg p$ if $(p,\bot)\in w$. A world $w$ is \emph{consistent} if there is no atom $p\in\mathbb{P}$ such that both $(p,\top)\in w$ and $(p,\bot)\in w$.
\end{definition}

Crucially, cognitive worlds record atomic determinations only; the evaluation of complex formulas is induced from them recursively, and modal structure is introduced separately in designated modal logics.

\bigskip
We designate two special worlds:
\begin{itemize}
\item[--] The \textit{shared knowledge} world, denoted $s$, a \emph{consistent} cognitive world representing universally accepted information. It serves as a lower bound: a world $w$ is \emph{admissible} iff $s \subseteq w$. The set $s$ may be empty or partial, but once determined, it specifies the space of admissible modal worlds.
\item[--] The \textit{reality} world, denoted $r$, another \emph{consistent} cognitive world providing the non-modal base for interpretation. The world $r$ must itself be admissible, i.e., it satisfies $s \subseteq r$.
\end{itemize}

\begin{definition}[Designated modal logic]\label{def:designated}
Let $r$ be the designated reality and $w$ an admissible world (i.e.\ $s\subseteq w$).
The \emph{designated modal logic} for $(r,w)$, written $\Sigma_{r,w}$, is the set of all formulas that are true at $r$ according to the evaluation clause
\[
r \models \Box \varphi \quad \text{iff} \quad w \models \varphi.
\]
% This clause is applied recursively at further modal levels. 
\end{definition}

% Since cognitive worlds assign values only to atoms, the truth of complex formulas is determined recursively from those atomic assignments, while modal formulas are evaluated by the designated pair $(r,w)$ through the clause above. Atoms and connectives are otherwise evaluated as in the non-bivalent base and assuming classical behavior for defined formulas.

According to this definition, cognitive worlds record atomic determinations only, while complex formulas inherit their values compositionally from those atomic assignments, according to the non-bivalent base and classical behavior wherever defined. In particular, the modal clause belongs only to the interaction between an arbitrary cognitive world $w$ and the designated reality $r$: thus, $\Box$ occurs only at that level, and greater modal depth arises only through further axiomatic assumptions, such as Axiom \textbf{4} or \textbf{5}.

As a notational convention, we write “$\Sigma_{r,w}\models \varphi$” to mean “$r\models \varphi$”, and “$\Sigma_{r,w}\models \Box\varphi$” to mean “$w\models \varphi$”.

This formulation abstracts away from syntactic derivability and focuses on the set-theoretic relationship between worlds. The class of logics so obtained is robust enough to accommodate standard modal axioms. As we will show, each such axiom corresponds to an inclusion constraint between $r$ and $w$ (e.g., $w\subseteq r$ for \textbf{T}, $r\subseteq w$ for \textbf{C}, and consistency of $w$ for \textbf{D}).

\begin{definition}[Cognitive classes]\label{def:classes}
Let $\mathbb{P}$ be a non-empty set of atoms, and let $\Sigma_{r,w}$ be as in definition~\ref{def:designated}. We define:
\begin{align}
B_r &:= \{ w \subseteq \mathbb{P}\times \{\top,\bot\} \mid s \subseteq w \}, \\
K_r &:= \{ w \subseteq \mathbb{P}\times \{\top,\bot\} \mid s \subseteq w \subseteq r \},\\ 
C_r &:= \{ w \subseteq \mathbb{P}\times \{\top,\bot\} \mid r \subseteq w \}.
\end{align}
\end{definition}

The classes $B_r$, $K_r$, and $C_r$ let us read modal constraints as \emph{set–theoretic inclusions} between valuation sets relative to the designated 
$s$ and $r$. For any admissible $w$, its position with respect to $s$ and $r$ determines exactly which axioms hold\footnote{Strictly speaking, $B_r$ depends only on $s$, not on $r$. We keep the index $r$ for notational symmetry with $K_r$ and $C_r$, and to emphasize that all three classes are defined relative to the same designated pair $(s,r)$.}. The following lemmas and the final inclusion theorem make this precise.

\begin{lemma}[K]\label{lemma1G}
If $s\subseteq w$ (i.e., $w$ is admissible), then the designated modal logic $\Sigma_{r,w}$ validates $\mathbf{K}$. If, moreover, $w$ is consistent, then $\Sigma_{r,w}$ validates $\mathbf{D}$ (hence $\mathbf{KD}$).
\end{lemma}

\begin{proof}
By definition~\ref{def:designated}, for every formula $\varphi$ we have $r \models \Box \varphi$ iff $w \models \varphi$.

To validate $\mathbf{K}$, assume $r\models \Box(\varphi\to\psi)$ and $r\models \Box\varphi$. Then $w\models \varphi\to\psi$ and $w\models \varphi$, hence $w\models \psi$, so $r\models \Box\psi$. Therefore $r\models \Box(\varphi\to\psi)\to(\Box\varphi\to\Box\psi)$.

To validate $\mathbf{D}$, assume $w$ is consistent and $r\models \Box\varphi$, hence $w\models \varphi$. Since defined formulas obey classical behavior, a consistent world cannot make both $\varphi$ and $\neg\varphi$ true. Thus if also $r\models \Box\neg\varphi$, then $w\models \neg\varphi$, contradicting the consistency of $w$. This concludes that $r\models \Box\varphi\to \neg\Box\neg\varphi$.
\end{proof}

\begin{lemma}[KT] \label{lemma2G}
If $s\subseteq w\subseteq r$, then the designated modal logic $\Sigma_{r,w}$ validates $\mathbf{KT}$.
\end{lemma}

\begin{proof}

By lemma~\ref{lemma1G}, $\Sigma_{r,w}$ validates $\mathbf{K}$. It remains to show $\mathbf{T}$, i.e., $r\models \Box\varphi\to \varphi$ for every formula $\varphi$. Assume $r\models \Box\varphi$. By definition~\ref{def:designated}, $r\models \Box\varphi$ iff $w\models \varphi$. Hence $w\models \varphi$. We prove $r\models \varphi$ by induction on $\varphi$.

\emph{Base case (atom).} Let $\varphi\equiv p$ and suppose $w\models p$. Then $(p,\top)\in w$. Since $w\subseteq r$, we have $(p,\top)\in r$, hence $r\models p$.

\emph{Inductive cases (connectives).} Propositional connectives are interpreted recursively from the truth of the immediate subformulas, according to the clauses of the non-bivalent base. By the induction hypothesis, each immediate subformula of $\varphi$ that is true at $w$ is also true at $r$. Therefore the same recursive clause that makes $\varphi$ true at $w$ makes it true at $r$ as well. Thus $r\models \varphi$.

Therefore $r\models \Box\varphi\to \varphi$, so $\Sigma_{r,w}$ validates $\mathbf{T}$, hence $\mathbf{KT}$.
\end{proof}

\begin{lemma}[KC] \label{lemma3G}
If $r\subseteq w$, then the designated modal logic $\Sigma_{r,w}$ validates $\mathbf{KC}$. If, moreover, $w$ is consistent, then $\Sigma_{r,w}$ validates $\mathbf{KDC}$.
\end{lemma}

\begin{proof}
By lemma~\ref{lemma1G}, $\Sigma_{r,w}$ validates $\mathbf{K}$. It remains to show $\mathbf{C}$, i.e., $r\models \varphi\to \Box\varphi$ for every formula $\varphi$. Assume $r\models \varphi$. We prove $w\models \varphi$ by induction on $\varphi$.

\emph{Base case (atom).} Let $\varphi\equiv p$ and suppose $r\models p$. Then $(p,\top)\in r$. Since $r\subseteq w$, we have $(p,\top)\in w$, hence $w\models p$.

\emph{Inductive cases (connectives).} Propositional connectives are interpreted recursively from the truth of the immediate subformulas, according to the clauses of the non-bivalent base. By the induction hypothesis, each immediate subformula of $\varphi$ that is true at $r$ is also true at $w$. Therefore the same recursive clause that makes $\varphi$ true at $r$ makes it true at $w$ as well. Thus $w\models \varphi$, and by definition~\ref{def:designated} we have $r\models \Box\varphi$.

Therefore $r\models \varphi\to \Box\varphi$, so $\Sigma_{r,w}$ validates $\mathbf{C}$.

Now assume that $w$ is consistent, and suppose $r\models \Box\varphi$. By definition~\ref{def:designated}, this holds iff $w\models \varphi$.
Since defined formulas obey classical behavior, a consistent world cannot make both $\varphi$ and $\neg\varphi$ true. Hence it cannot be that $w\models \neg\varphi$, and therefore it cannot be that $r\models \Box\neg\varphi$.
Therefore $r\models \Box\varphi\to \neg\Box\neg\varphi$, and $\Sigma_{r,w}$ validates $\mathbf{KDC}$.
\end{proof}

\begin{theorem}[Inclusion Theorem]\label{thm:inclusion-classes}
Let $\Sigma_{r,w}$ be as in definition~\ref{def:designated}, and $B_r,K_r,C_r$ as in definition~\ref{def:classes}. Then for any subset $x\subseteq \mathbb{P}\times\{\top,\bot\}$:
\begin{itemize}\setlength\itemsep{-3pt}
\item[(i)] $\ x\in B_r \Longleftrightarrow \Sigma_{r,x}\ \text{is defined and validates }\mathbf{K}$;
\item[(ii)] $\ x\in K_r \Longleftrightarrow \Sigma_{r,x}\ \text{is defined and validates }\mathbf{KT}$;
\item[(iii)] $\ x\in C_r \Longleftrightarrow \Sigma_{r,x}\ \text{is defined and validates }\mathbf{KC}$;
\item[(iv)] $\ x\in B_r \text{ and is consistent} \Longleftrightarrow \Sigma_{r,x}\ \text{is defined and validates }\mathbf{KD}$;
\item[(v)] $\ x\in C_r \text{ and is consistent} \Longleftrightarrow \Sigma_{r,x}\ \text{is defined and validates }\mathbf{KDC}$.
\end{itemize}
\end{theorem}

\begin{proof}
Let $x\subseteq \mathbb{P}\times\{\top,\bot\}$ and assume that whenever $\Sigma_{r,x}$ is used, $x$ is admissible, i.e.\ $s\subseteq x$ (by definition~\ref{def:designated}).

\smallskip
\noindent\emph{(i)} If $x\in B_r$, then $s\subseteq x$, hence $\Sigma_{r,x}$ is defined; by lemma~\ref{lemma1G}, axiom $\mathbf{K}$ is valid at $r$, therefore $\Sigma_{r,x}$ validates $\mathbf{K}$. Conversely, $\Sigma_{r,x}$ is defined, hence $s\subseteq x$ and $x\in B_r$.

\smallskip
\noindent\emph{(ii)} If $x\in K_r$, then $s\subseteq x\subseteq r$; by lemma~\ref{lemma1G}, $\mathbf{K}$ is valid at $r$, and by lemma~\ref{lemma2G}, $\mathbf{T}$ is valid at $r$, so $\Sigma_{r,x}$ validates $\mathbf{KT}$. Conversely, if $\Sigma_{r,x}$ validates $\mathbf{KT}$, then in particular $\mathbf{T}$ holds at $r$. For any $(p,\top)\in x$, we have $x\models p$, hence $\Sigma_{r,x}\models \Box p$ and so $r\models \Box p$; by $\mathbf{T}$, $r\models p$, thus $(p,\top)\in r$. The case $(p,\bot)\in x$ is analogous. Hence $x\subseteq r$. Since $\Sigma_{r,x}$ is defined, $s\subseteq x$, so $x\in K_r$.

\smallskip
\noindent\emph{(iii)} If $x\in C_r$, then $r\subseteq x$; by lemma~\ref{lemma3G}, $\mathbf{C}$ is valid at $r$, so $\Sigma_{r,x}$ validates $\mathbf{KC}$. Conversely, if $\Sigma_{r,x}$ validates $\mathbf{KC}$, then in particular $\mathbf{C}$ holds at $r$. For any $(p,\top)\in r$, we have $r\models p$, hence by definition~\ref{def:designated} $r\models \Box p$; therefore $\Sigma_{r,x}\models \Box p$ and $x\models p$, so $(p,\top)\in x$. The case $(p,\bot)\in r$ is analogous. Thus $r\subseteq x$, i.e.\ $x\in C_r$.

\smallskip
\noindent\emph{(iv)} If $x\in B_r$ and $x$ is consistent, lemma~\ref{lemma1G} yields that $\mathbf{D}$ is valid at $r$, hence $\Sigma_{r,x}$ validates $\mathbf{KD}$. Conversely, if $\Sigma_{r,x}$ validates $\mathbf{KD}$, then for every $\varphi$ we have $r\models \Box\varphi \to \neg \Box\neg\varphi$. If $x$ were inconsistent, there would be $p$ with $x\models p$ and $x\models \neg p$, hence $r\models \Box p$ and $r\models \Box\neg p$, contradicting $\mathbf{D}$. Therefore $x$ is consistent and, since $\Sigma_{r,x}$ is defined, $x\in B_r$.

\smallskip
\noindent\emph{(v)} If $x\in C_r$ and $x$ is consistent, lemma~\ref{lemma3G} gives that $\mathbf{C}$ and $\mathbf{D}$ are valid at $r$, hence $\Sigma_{r,x}$ validates $\mathbf{KDC}$. Conversely, if $\Sigma_{r,x}$ validates $\mathbf{KDC}$, then $\mathbf{C}$ holds at $r$, which by (iii) implies $r\subseteq x$ (so $x\in C_r$), and $\mathbf{D}$ holds at $r$, which by (iv) implies that $x$ is consistent.
\end{proof}

Theorem~\ref{thm:inclusion-classes} proves its result at the level of atomic determinations. Additional modal principles such as \textbf{4} and \textbf{5} may constrain the iterated behavior of the modality, but they do not change the underlying set-theoretic relations between the atomic contents of $r$ and $w$. For this reason, the inclusion constraints associated with \textbf{K}, \textbf{T}, \textbf{C}, and \textbf{D} remain the relevant ones for the classes defined below.

The Inclusion Theorem thus turns proof–theoretic strength into a \emph{geometric} picture: any world $w$ with $s \subseteq w$ is already doxastic ($\mathbf{K}$, $\mathbf{K45}$, $\mathbf{KD}$, $\mathbf{KD45}$); moving $w$ \emph{toward} $r$ yields epistemic strength ($\mathbf{KT}$, $\mathbf{KT45}$), moving \emph{beyond} $r$ yields conjectural strength ($\mathbf{KC}$, $\mathbf{KDC}$). Because the correspondences are stated at the atomic, set–theoretic level, they remain stable over the non-bivalent base and require no additional frame postulates.

\subsection{Cognitive Worlds}\label{subsec:cognitiveworlds}

The structure established by the inclusion theorem allows us to provide a first classification of all  worlds satisfying at least \textbf{K}, based on their relationship with the reality world $r$ and the shared knowledge $s$. Each world $w$ defines a possible configuration of beliefs, and its relationship with $r$ determines which axioms a logic satisfies and, more profoundly, what type of belief it represents. It is thus possible to identify at least six well-identified categories:

\begin{itemize}
    \item \textbf{Reality}: the boundary case $w=r$. In addition to being the logic of boxless predicates, by theorem~\ref{thm:inclusion-classes}, both $\mathbf{T}$ and $\mathbf{C}$ hold at $r$, so reality belongs to the intersection of the epistemic and conjectural classes. 

    \item \textbf{Doxastic World}: any world $w$ such that $s \subseteq w$. Doxastic requirements hold toward systems such as $\mathbf{K}$ and $\mathbf{K45}$. If consistency is maintained we obtain $\mathbf{KD}$ and $\mathbf{KD45}$. 

    \item \textbf{Epistemic World}: any world $w$ such that $w \subseteq r$. Since truth at $w$ guarantees truth at $r$, axiom $\mathbf{T}$ holds. Thus, epistemic worlds correspond to systems like $\mathbf{KT}$ and $\mathbf{KT45}$.

    \item \textbf{Conjectural World}: any world such that $r \subseteq w$. An extension of reality $r$ possibly assigning values to propositions undefined in $r$. This represents extended or speculative knowledge, not delusional but consistent with known reality. These systems are characteristic of logics such as $\mathbf{KC}$ and $\mathbf{KDC}$.

    \item \textbf{Delusional World}: any world $w$ for which there exists $p\in\mathbb{P}$ with either $(p,\top)\in w$ and $(p,\bot)\in r$, or $(p,\bot)\in w$ and $(p,\top)\in r$. In short, $w$ contradicts $r$ on some atom; this corresponds, obviously, to a false belief in its cognitive world. In general, such worlds satisfy neither Axiom $\mathbf{T}$ nor Axiom $\mathbf{C}$.

    \item \textbf{Opinion World}: any world $w$ whose assigned atoms are disjoint from those of $r$ beyond the shared $s$, i.e. $\mathrm{dom}(w)\cap \mathrm{dom}(r)=\mathrm{dom}(s)$.\footnote{where $\mathrm{dom}(x):=\{\,p\in\mathbb{P}\mid (p,\top)\in x \text{ or } (p,\bot)\in x\,\}$} These worlds are disconnected from reality; in general, like delusions, such worlds do not satisfy principles stronger than Axiom $\mathbf{K}$.
\end{itemize}

This classification, although partial, provides interesting ways to represent a complex set of interrelations between reality and subjective cognitive states, and allows for newer types of cognitive logics to arise in a controlled fashion.

\paragraph{Example.}
Let $\mathbb{P}=\{a,b,c,d,e,f\}$. For readability, we write $p$ for $(p,\top)$ and 
$¬p$ for $(p,\bot)$. Let $s=\{a,\neg b\}$ and $r=\{a,\neg b,\neg c,d\}$.
\begin{itemize}
\item \textbf{Doxastic worlds}: all $w\supseteq s$.
\item \textbf{Epistemic worlds}: all $w$ with $s\subseteq w\subseteq r$, i.e.
$\{a,\neg b\}$, $\{a,\neg b,\neg c\}$, $\{a,\neg b,d\}$, $\{a,\neg b,\neg c,d\}$.
\item \textbf{Conjectural worlds}: all $w\supseteq r$, e.g.
$\{a,\neg b,\neg c,d\}$, $\{a,\neg b,\neg c,d,e\}$, $\{a,\neg b,\neg c,d,\neg e\}$, etc. 

\item \textbf{Delusional worlds}: any $w$ contradicting $r$ on some atom, e.g.
$\{a,\neg b,c,d\}$, $\{a,\neg b,\neg c,\neg d\}$, etc.
\item \textbf{Opinion worlds}: any $w$ sharing no atoms with $r$ except the shared ones, e.g.
$\{a,\neg b,e\}$, $\{a,\neg b,e,\neg f\}$, etc. 
\end{itemize}
For instance, with $r$ as above the world $w=\{a,\neg b,\neg c,d,\neg e\}$, validates the modal logic
$a,\neg b,\neg c,d,\Box a,\Box\neg b,\Box\neg c,\Box d,\Box\neg e$.

\section{Base logics for conjectural logics}\label{sec:baselogics}

Up to this point, conjectural logic has been developed over an abstract non-bivalent base logic, specified only through the general constraints introduced earlier. At that level, the framework does not depend on any particular choice of truth values, semantic clauses, or knowledge structures. Indeed, its central requirements are more general: there must be a settled base, there must be room for further determinations beyond that base, and those further determinations must be able to generate distinct yet structurally related worlds.

Several established logical systems instantiate exactly this pattern in concrete and recognizable ways. Three cases are discussed here: Supervaluation Semantics, Weak Kleene logic, and Description Logic. They can be considered natural semantic settings for, e.g., disagreement over incompatible completions of a shared base, future contingents and not-yet-defined propositions, or more precisely for open hypotheses over a shared knowledge base.

\subsection{Supervaluation Semantics}\label{subsec:supervaluation}

Supervaluation semantics \cite{Fine1975Supervaluation} starts from a partial assignment of classical truth values and interprets indeterminacy through the total valuations extending that assignment. This gives a standard semantic treatment of partial information: what is already determined remains determined, while what is still open may be completed in more than one admissible way.

Let $\mathsf{At}$ be a set of atomic propositions. A \emph{partial valuation} is a partial function
\[
V:\mathsf{At}\rightharpoonup \{\top, \bot\}.
\]
This is the standard supervaluation instance, and it is the same kind of partial assignment introduced abstractly in Section~4: the domain of $V$ determines the settled fragment of the factual base, while atoms outside $\mathrm{dom}(V)$ remain not yet determined.

\begin{definition}[Completion]
A \emph{completion} of $V$ is a total classical valuation
\[
W:\mathsf{At}\to \{\top, \bot\}
\]
such that, for every $p\in \mathrm{dom}(V)$,
\[
W(p)=V(p).
\]
The set of all completions of $V$ may be denoted by $\mathrm{Comp}(V)$.
\end{definition}

Thus, each completion preserves the assignments already determined in $V$ and extends only the atoms on which $V$ was undefined.

\begin{definition}[Supertruth and superfalsity]
Given a formula $\varphi$:
\begin{itemize}
\item $\varphi$ is \emph{supertrue} at $V$ iff, for every $W\in \mathrm{Comp}(V)$, $W(\varphi)=\top$
\item $\varphi$ is \emph{superfalse} at $V$ iff, for every $W\in \mathrm{Comp}(V)$, $W(\varphi)=\bot$
\end{itemize}
If neither condition holds, $\varphi$ is undetermined at $V$.
\end{definition}

So the basic semantic pattern is the following:
\begin{itemize}
\item $V$ determines the common settled base
\item $\mathrm{Comp}(V)$ determines the admissible extensions of that base
\item truth at the partial level is obtained by quantifying over all such extensions
\end{itemize}

This gives a direct bridge to the abstract framework of Section~4. There, the factual base was introduced in general form as a partial assignment establishing some content while leaving other content open. Here, that structure is realized in a standard and fully explicit way: the factual base is the partial valuation $V$ itself, and the accessible semantic developments are its completions. Nothing already settled is revised: only what was not yet determined may vary across worlds.

Thus conjectural accessibility can be read in supervaluation semantics as movement across incompatible but conservative continuations of the same initial base, and different accessible worlds agree on the shared settled core and differ in the determination of the still open parts. 
This yields a natural model of disagreement when represented as different admissible completions of the same underlying partial base. 

\subsection{Weak Kleene Logic}\label{subsec:WeakKleene}

Weak Kleene logic \cite{Kleene1952Metamathematics} is a three-valued logic whose truth values are $\top$, $\bot$, and $u$ for \textit{true}, \textit{false} and \textit{undefined} respectively.

Its central feature is the propagation of undefinedness. When the evaluation of a compound formula depends on a component that is undefined, the compound formula also becomes undefined. This gives Weak Kleene logic a semantics of genuine partiality: undefinedness is not an external lack of information, but an explicit internal value.

Thus the valuation in this setting is a total function of the form
\[
V:\mathsf{At}\to \{\top,\bot,u\}.
\]

The basic clauses for the connectives may be summarized as follows:
\begin{itemize}
    \item \textit{Negation}: $\neg \top = \bot$, $\neg \bot = \top$, $\neg u = u$;
    \item \textit{Conjunction}: $\varphi \wedge \psi$ is undefined whenever neither conjunct determines the result and one relevant component is undefined;
    \item \textit{Disjunction}: $\varphi \vee \psi$ is undefined whenever neither disjunct determines the result and one relevant component is undefined;
    \item \textit{Implication}: defined as $\varphi \rightarrow \psi := \neg \varphi \vee \psi$, inherits the same contagious behavior of undefinedness of its components.
\end{itemize}

Differently from Strong Kleene semantics, where defined values may in some cases determine the result despite the presence of an undefined component, in Weak Kleene logic undefinedness spreads more aggressively whenever it enters into the effective evaluation of the formula. This makes the logic especially appropriate when undefinedness is meant to represent a lack of determination.

Several structural features are important in the present context:
\begin{itemize}
\item \emph{partiality}: not every proposition must already have a classical value
\item \emph{undefinedness propagation}: compounds depending on undefined components may remain undefined
\item \emph{block of classical tautologies}: formulas such as $\varphi \vee \neg \varphi$ are not generally valid
\item \emph{monotonicity}: adding determinate information does not destroy earlier determinate evaluations
\item \emph{paracompleteness}: excluded middle is not valid in general
\item \emph{non-triviality}: $\top$ and $\bot$ remain distinct truth values with ordinary logical significance
\end{itemize}

This makes Weak Kleene logic well suited to informational states in which some claims are still not yet determined. A present world may already settle part of the propositional base while leaving another part undefined. Future contingents provide a natural example: a proposition concerning a future event may fail to be true and fail to be false at the current stage, precisely because the present state does not yet determine it.

Weak Kleene's actual undefinedness supports the conjectural extension when understood as branching from the undefined region of a world toward later or alternative determinations. What is already assigned $\top$ or $\bot$ remains part of the common base, while what is assigned $u$ remains open to further specification across accessible worlds. The transition from undefinedness to later determination is therefore the natural connection between these logics. 

\subsection{Description Logic}\label{subsec:Description}

Description Logic \cite{BaaderEtAl2003DLHandbook} is a family of formalisms for representing structured knowledge. Its language is built from three main kinds of symbols:
\begin{itemize}
\item \emph{concepts}: unary predicates, used to classify objects
\item \emph{roles}: binary predicates, used to represent relations
\item \emph{individuals}: constants naming elements of the domain
\end{itemize}

A Description Logic knowledge base is usually divided into two components:
\begin{itemize}
\item \emph{TBox}: the terminological component, containing concept inclusions and other general constraints
\item \emph{ABox}: the assertional component, containing facts about individuals
\end{itemize}

A minimal semantic presentation is also standard. An interpretation $\mathcal{I}$ consists of a non-empty domain $\Delta^{\mathcal{I}}$ and an assignment mapping:
\begin{itemize}
\item each concept $C$ to a subset $C^{\mathcal{I}}\subseteq \Delta^{\mathcal{I}}$
\item each role $r$ to a binary relation $r^{\mathcal{I}}\subseteq \Delta^{\mathcal{I}}\times \Delta^{\mathcal{I}}$
\item each individual $a$ to an element $a^{\mathcal{I}}\in \Delta^{\mathcal{I}}$
\end{itemize}

At the level of satisfaction, this framework is fundamentally bivalent: an assertion or inclusion is either satisfied in an interpretation or it is not. Yet, in Description Logic a knowledge base may determine only part of the relevant informational structure: a shared ontology and a shared body of facts may already be established, while further classifications, assertions, or constraints remain open.

Therefore, Description Logic becomes a natural base for conjectural extension: a family of worlds may share a common TBox and a common factual ABox core, so that they agree on the conceptual vocabulary, on the general ontological structure, and on a settled fragment of assertions about individuals. Conjectural development arises when additional statements are added locally on top of the shared base, with new concept memberships, new role assertions, further terminological commitments, or more specific classificatory extensions.

Each different extension remains coherent with the shared base while possibly becoming mutually incompatible with the others. Accessible conjectural worlds are thus understood as structured continuations obtained by adding further hypotheses beyond that base. 

\section{Settlements and dynamic conjectural logics}\label{sec:settlements}
This section turns from the static framework developed so far to its dynamic counterpart. The question is how to represent transitions that affect cognitively structured states and, in particular, how to describe the moment at which a possibility that was previously only conjectured acquires the status of current reality.

In the previous sections, we introduced modal logics that describe different ways in which an agent or a community may stand in relation to reality: by believing it, knowing it, or conjecturing it. We grouped these systems under the umbrella of \textit{cognitive modal logics}. A natural next step is to ask whether there is a corresponding family of cognitive dynamic logics that studies the transformation of such states and not only their structure. Dynamic doxastic logics describe the evolution of beliefs, and dynamic epistemic logics describe the evolution of information or knowledge. The purpose of this section is to identify the dynamic counterpart of conjectural logic.

For the conjectural case, the relevant dynamic problem is the passage by which a possibility that was previously only conjectured acquires the status of current reality. This motivates the introduction of a dynamic logic built around a new operator called \textit{settlement}. If conjectural logic describes a space of possible extensions above a reality that is still only partially determined, then a dynamic conjectural logic must be able to settle an extension, that is, to describe the transition by which one of those extensions is henceforth treated as real.

Dynamic logic becomes relevant as soon as one is interested not only in the structure of a cognitive state, but also in the transitions by which such a state changes under suitable events. In the present context, this gives rise to what may be called \emph{cognitive dynamic logics}: dynamic logics whose states are cognitively characterized, for instance as doxastic, epistemic, or conjectural. Their shared feature is a semantic role rather than a single technical format, since they describe how events transform states that are interpreted in cognitive terms.

In dynamic doxastic logic \cite{vanbenthem2007dynamic} the states under consideration are belief states, often represented by plausibility orderings, preference structures, or other devices meant to encode what an agent takes to be more or less credible. The characteristic operators of this setting modify the doxastic organization of the space itself: they may reorder plausibilities, revise previous commitments, or determine which worlds count as doxastically preferred after a change. In this sense, dynamic doxastic logic studies the evolution of belief rather than the accumulation of truths, and is therefore naturally related to the broader tradition of belief revision and priority change.

Dynamic epistemic logic \cite{Plaza1989PublicAnnouncements} addresses a second, equally central case. Here the evolving states are epistemic or informational states, and the characteristic events are announcements, observations, communications, and more general epistemic actions. The best known examples are public announcement logic and event-model based updates, where the effect of an event is to modify what agents know, what they can distinguish, or which worlds remain epistemically accessible after the event. What matters especially for the present discussion is that these dynamic operators typically act on the modal and informational side of the model, while the non-modal reality in the background is usually treated as already established and fully determined.

The conjectural case raises a different problem. If conjectural logic describes a situation in which reality itself is only partially determined and several conjectural extensions are already available above it, then the corresponding dynamic logic should represent the transition by which one of these extensions becomes current reality. The role played below by settlements is precisely to model this transition, thereby extending to the conjectural case the same general perspective that dynamic doxastic and dynamic epistemic logics apply to belief and knowledge.

The dynamics considered here is defined on models with a designated reality. The relevant semantic object is therefore the pair
$(\mathcal M,r)$,
where
$\mathcal M=\langle W,R,\mathcal V\rangle$
is a Kripke model and $r\in W$ is the designated reality.

A settlement concerns an atom that is still undetermined at $r$. In that situation, the conjectural structure above $r$ contains two opposed extensions: one in which the atom is made true, and one in which the contrary commitment is made true instead. The formal role of settlements is to turn one of these already available extensions into the new designated reality.
\begin{definition}
Let $(\mathcal M,r)$ be a model with designated reality, and let $p$ be an atom. A world $u\in W$ is the \textbf{least definedness-preserving extension} of $r$ for $p$ if:
\begin{enumerate}
\item $u$ is a definedness-preserving extension of $r$ (as in definition~\ref{def:definedness-preserving-extension});
\item if $V_r(p)$ is defined, then $u=r$;
\item if $V_r(p)$ is undefined, then $V_u(p)=\top$;
\item for every world $w\in W$, if $w$ is a definedness-preserving extension of $r$ and $V_w(p)=\top$, then
\[
V_u \subseteq V_w.
\]
\end{enumerate}
\end{definition}

We write $r^p$ for the least definedness-preserving extension of $r$ in which $p$ is true. This choice is purely conventional: when a new determination is introduced, one may equivalently consider the opposite extension in which $p$ is false. For simplicity, we write the positive case as $r^p$. If the opposite determination is needed, let $q:=\neg p$ and write $r^q$ for the least definedness-preserving extension in which $q$ is true, that is, the extension in which $p$ is false.

\begin{definition}\label{def:settlement}
A \textbf{settlement} is the transition
\[
(\mathcal M,r)\xrightarrow{\mathrm{settle}(p)}(\mathcal M,r^p),
\]
where the designated reality changes from $r$ to $r^p$, while the underlying Kripke model $\mathcal M$ remains the same.
\end{definition}

Some immediate consequences can already be read off from this definition.

\begin{proposition}\label{prop:settlement-cognitive-worlds}
Let $(\mathcal M,r)$ be a model with designated reality, and let $r^p$ be the least definedness-preserving extension of $r$ for $p$. Then:

\begin{enumerate}
\item relative to the new designated reality $r^p$, the previous reality $r$ is an epistemic world;
\item if $q:=\neg p$, then the opposite extension $r^q$ is delusional relative to the new designated reality $r^p$;
\item every world $w$ such that $r^p \subseteq w$ is conjectural relative to the new designated reality $r^p$.
\end{enumerate}
\end{proposition}

\begin{proof}
(1) A world is epistemic relative to a designated reality $r^p$ when it is included in $r^p$ and still contains the shared base $s$. Since $r$ is admissible, we have $s \subseteq r$, and by (1) we also have $r \subseteq r^p$. Therefore
$s \subseteq r \subseteq r^p$,
so $r$ is epistemic relative to the new designated reality.

(2) Let $q:=\neg p$. By construction, $r^p$ is the extension in which $p$ is determined positively, whereas $r^q$ is the extension in which the opposite determination is taken. Hence $r^p$ and $r^q$ disagree on the value of $p$: one contains $(p,\top)$ and the other contains $(p,\bot)$. Therefore $r^q$ contradicts $r^p$ on at least one atom, and so $r^q$ is delusional relative to $r^p$.

(3) By definition, a world is conjectural relative to a designated reality precisely when it extends that reality. Thus every world $w$ such that
$r^p \subseteq w$
is conjectural relative to the new designated reality $r^p$.
\end{proof}

\bigskip
Settlements support progressive determination of reality: they can be used to describe not only an isolated transition, but a process in which further undetermined content is asserted step by step.

\begin{definition}
Let
\[
\Sigma = [\mathrm{settle}(p_1),\dots,\mathrm{settle}(p_n)]
\]
be a finite sequence of settlements. We define recursively
\[
(\mathcal M,r_0):=(\mathcal M,r),
\qquad
(\mathcal M,r_{i+1}) := (\mathcal M,r_i^{p_{i+1}})
\]
whenever each least definedness-preserving extension $r_i^{p_{i+1}}$ exists.
\end{definition}

\begin{proposition}[Iterability of settlements]
If the sequence $\Sigma = [\mathrm{settle}(p_1),\dots,\mathrm{settle}(p_n)]$ is well-defined, then
\[
V_{r_0}\subsetneq V_{r_1}\subsetneq \cdots \subsetneq V_{r_n}.
\]
\end{proposition}

\begin{proof}
By definition, each $r_{i+1}=r_i^{p_{i+1}}$ is a least definedness-preserving extension of $r_i$ in which $p_{i+1}$ is true. Hence
\[
V_{r_i}\subseteq V_{r_{i+1}}.
\]
Moreover, $V_{r_i}(p_{i+1})$ is undefined before the corresponding settlement, while $V_{r_{i+1}}(p_{i+1})=\top$. Therefore
\[
V_{r_i}\subsetneq V_{r_{i+1}}.
\]
Repeating this argument at every step yields the whole chain.
\end{proof}

A second structural point concerns the order of determination. When two settlements are compatible, changing their order does not affect the resulting factual base.

\begin{theorem}[Commutation of compatible settlements]
Let $p$ and $q$ be atoms such that $V_r(p)$ is undefined, $V_r(q)$ is undefined, and the iterated least definedness-preserving extensions $(r^p)^q$ and $(r^q)^p$ both exist. Then
\[
V_{(r^p)^q}=V_{(r^q)^p}.
\]
In particular, if worlds are identified by their factual bases, the two iterated settlements lead to the same resulting reality.
\end{theorem}

\begin{proof}
Since $r^p$ is the least definedness-preserving extension of $r$ in which $p$ is true, every world extending $r$ and making both $p$ and $q$ true must extend $r^p$. Hence $(r^p)^q$ is minimal among worlds extending $r$ and making both $p$ and $q$ true.

By the symmetric argument, $(r^q)^p$ is also minimal among worlds extending $r$ and making both $p$ and $q$ true.

Therefore $V_{(r^p)^q}\subseteq V_{(r^q)^p}$ and, symmetrically,
\[
V_{(r^q)^p}\subseteq V_{(r^p)^q}.
\]
Hence
\[
V_{(r^p)^q}=V_{(r^q)^p}.
\]
\end{proof}

Together, these properties yield a structured dynamics of progressive determination.

\bigskip

Once settlements have been introduced, they make it possible to define a genuine dynamic extension of the static conjectural framework developed in the previous sections. We call this extension \emph{Conjectural Dynamic Logic} (\emph{CDL}). CDL should not be understood as a single isolated formal system, but as the dynamic layer corresponding to the family of conjectural logics already defined above. Its general static basis is therefore $\mathbf{KC}$, while $\mathbf{KDC}$ is the more specific case in which conjectural states are internally coherent. The point of CDL is to add to that static framework the resources needed to represent transitions by which a partially determined reality becomes progressively more determined.

The need for such an extension follows directly from the semantic role of conjectural worlds. In the static setting, the designated reality does not necessarily exhaust what is semantically available in the model. Above it there may already exist several definedness-preserving extensions, each of which develops the current reality in a different admissible direction. As long as these extensions remain merely conjectural, the logic describes their order, their compatibility with the actual base, and the way in which truths propagate from the designated reality to its extensions. What the static setting does not yet express on its own is the moment at which one of those extensions acquires the status of current reality. CDL is introduced precisely in order to represent that passage.

The language of CDL is obtained by extending the language of the underlying conjectural logic with dynamic modalities corresponding to settlements. Thus, if $\varphi$ is a formula of the static language and $p$ is an atom, then
$[\mathrm{settle}(p)] \varphi$
is a formula of the dynamic language. The intended reading is that $\varphi$ holds after the settlement of $p$, once the designated reality has moved to the extension in which $p$ is settled.

Every atom that is still undetermined at the designated reality gives rise to two opposed conjectural directions: one in which the atom is made true, and one in which the corresponding contrary commitment is made true instead. This is why the notation $r^p$ is meaningful only against the background of a model that already contains conjectural extensions above $r$. The settlement operator acts on a structure in which they are already present and semantically ordered. For the same reason, if $p$ is already true at $r$, then settling $p$ leaves the designated reality unchanged, so that $r^p=r$. Symmetrically, if the contrary positive commitment is already true at $r$, its settlement is equally idle. A settlement that would impose the opposite value on an atom that is already determined is excluded, because the function of settlements is to determine what is still open, not to move from one fixed reality to another incompatible one.

The semantic clause for the dynamic modality is therefore:
\[
(\mathcal M,r)\models [\mathrm{settle}(p)]\varphi
\quad\text{iff}\quad
(\mathcal M,r^p)\models \varphi.
\]
The clause evaluates $\varphi$ at the world obtained by taking the least definedness-preserving extension of $r$ in which $p$ is true and promoting that extension to actuality. The underlying model remains fixed; the change concerns the designated reality.

CDL therefore represents transitions internal to an established conjectural space. Dynamic development consists in the progressive determination of a reality that was already incomplete. The structure of conjectural worlds provides the possible extensions in advance, and settlements select one of them as newly current. In this way, CDL preserves full continuity with the static framework: the same model that represents an incomplete reality together with its conjectural extensions also supports the transitions by which those extensions become actual in turn.

The properties established earlier in this section clarify why this deserves to count as a logic and not merely as an isolated operator. Settlements are iterable, so reality can become more determined through a finite sequence of transitions. They also commute whenever the relevant determinations are compatible, so that independent acts of determination lead to the same factual base regardless of order. These two facts show that CDL is not limited to representing one-step updates. It can describe a structured process in which an initially partial reality develops through successive settlements while preserving the semantic organization of its conjectural extensions.

CDL is the dynamic completion of the conjectural framework introduced earlier in the paper. Static conjectural logic describes the ordered space of extensions above a partially determined reality. CDL adds the means to represent the passage by which one such extension becomes designated reality and may in turn serve as the base for further settlements. The result is a dynamic logic whose operators are tailored to the semantic profile of conjectural states themselves. For this reason, CDL provides a natural and internally motivated dynamic counterpart to conjectural logic.

The main thesis of this section is that dynamic operators are constrained by the semantic character of the static logic on which they are built. Once the underlying states are interpreted cognitively, the form of the dynamic transition cannot be chosen independently of what those states already are. The operator has to answer the kind of openness, stability, or incompleteness that the static framework makes possible.

At the broadest level considered in this paper, in line with the general perspective adopted throughout the paper, all the cognitive states under discussion can be treated as doxastic, because they represent local commitments of an actor or a community. This is why the minimal common modal background is $\mathbf K$. Doxastic dynamic logic therefore occupies the most general level: it studies how such commitments change. Dynamic epistemic logics describe a more specific situation, since they concern cognitive states supported by knowledge or information. Their characteristic operators, such as public announcement, observation, and communication, are appropriate when the dynamic problem concerns informational accessibility and epistemic discrimination, typically against the background of a reality that is already taken as determined.

CDL answers a different semantic situation. In conjectural logic, the static base already contains a reality that is only partially determined together with conjectural extensions above it. For this reason, settlement is the fitting dynamic operator. It addresses the passage by which one conjectural possibility acquires the status of designated reality. The operator therefore follows from the semantic shape of the static space itself: where the base already contains ordered extensions above an incomplete reality, dynamic progress consists in promoting one of them to actuality.

AGM belief revision is a natural comparison because it also studies cognitively meaningful change under new information. Its classical operations are expansion, contraction, and revision, as introduced in the standard AGM framework \cite{AlchourronGardernforsMakinson1985AGM}. But AGM works on belief sets that must be reorganized, whereas CDL models the transition by which a conjectural extension becomes current reality inside a semantically structured space. The contrast concerns the target of the operation: in one case the organization of accepted commitments, in the other the designation of one extension within a partially determined reality.

It is useful for representing situations in which a conjecture becomes true, or, in a philosophically different but formally analogous way, situations in which a community accepts a previously conjectural result as true. The effective victory of team A in today’s match has exactly this form, because one branch of a previously open structure becomes actual. The attribution of a painting to a specific painter can take the same form when certainty is not available and competing hypotheses are instead judged less plausible. CDL gives a formal account of that transition, and for this reason it completes the static theory of conjectural logics with a corresponding dynamic layer.

\section{Conjectures and related logical frameworks} \label{sec:related}

Conjectural reasoning lies close to several well established logical traditions. This proximity is not accidental: whenever reasoning proceeds beyond what is already given, logic must account for the relation between an initial base and the conclusions, commitments, or alternative states that arise from it.

For that reason, a number of existing frameworks naturally come into view. Four of them are especially relevant here: Default Logic, Autoepistemic Logic, AGM Belief Revision, and Standpoint Logic. Each captures a genuine phenomenon that stands near conjectural reasoning, even if none of them is centered on conjecture in the sense developed in this paper.

\paragraph{Default Logic} is a standard framework for nonmonotonic reasoning \cite{Reiter1980Default}. It starts from a background theory and enriches it by means of default rules, namely inference patterns that license conclusions in the absence of defeating information. The background theory provides the explicitly accepted content, while the default rules allow the system to go further whenever consistency permits it.

The resulting process may generate one or more extensions. Each extension represents a possible body of conclusions obtained by applying the available defaults to the background theory. The central phenomenon modeled in this way is defeasible inference: conclusions are accepted not because they are deductively forced by the theory alone, but because they remain admissible as long as no contrary information blocks them.

\paragraph{Autoepistemic Logic} is a modal framework for reasoning about an agent's own beliefs \cite{Moore1985Autoepistemic}. Its characteristic feature is the use of a modal operator through which the agent may represent what is believed, and, symmetrically, what is not believed. The logic is therefore concerned with reflection on one's own epistemic or doxastic condition.

Its central constructions are stable expansions, or stable belief states. These are sets of formulas closed not only under ordinary consequence, but also under the reflective conditions generated by the agent's doxastic commitments. In this way, Autoepistemic Logic models introspection and belief reflection: the agent reasons from what it takes itself to believe, and from what it does not take itself to believe, producing a structured account of self-referential doxastic reasoning.

\paragraph{AGM Belief Revision} studies the rational modification of accepted beliefs under the impact of incoming information \cite{AlchourronGardernforsMakinson1985AGM}. Its central concern is theory change: given a belief set, or more generally a theory, how should that set be altered when new information arrives and must be incorporated in a rational way.

The classical AGM framework distinguishes three basic operations: expansion, contraction, and revision. Expansion adds new information without giving up previous beliefs; contraction removes a belief or commitment; revision incorporates incoming information while restoring overall consistency when needed. The framework is thus designed to analyze how a belief set changes over time under rational pressure, and how such change may be constrained by principles of minimal disturbance and coherence.

\paragraph{\textit{Ceteris Paribus} Logics} are designed to compare situations under conditions of restricted variation \cite{vanBenthemGirardRoy2009CeterisParibus}. Their central idea is that some relevant part of the background must be kept fixed, while selected differences are allowed to vary. In this way, they provide a formal treatment of all-else-being-equal comparison: one asks what follows, or what changes, when certain factors are modified against a stable contextual frame. The resulting framework is therefore naturally suited to comparative reasoning, where the logical focus falls on the significance of controlled variation rather than on the generation of new commitments from an incomplete factual base.

\paragraph{Standpoint Logic} is concerned with the representation of knowledge or assertions from multiple standpoints or perspectives \cite{GomezAlvarezRudolph2022Standpoint}. It is particularly suited to situations of semantic plurality or heterogeneity, where the same domain may be organized, described, or assessed differently depending on the standpoint from which it is considered. The framework is therefore sensitive to the fact that representation is often perspective-dependent rather than globally uniform.

Within this setting, different standpoints may support different acceptable interpretations or precisifications of the same subject matter. Viewpoint-dependent commitments are not treated as mere noise around a single fixed description, but as part of the structure through which complex domains are articulated. Standpoint Logic is thus naturally relevant whenever one wishes to model coexistence among distinct yet internally organized perspectives on the same domain.

\bigskip

Although there are several similarities with what these logics represent, the notion of conjecture developed in this paper is different from them and more specific: a conjecture is a cognitive commitment based on a shared factual base. That base is partial, embedded in a non-bivalent setting, and conjectural scenarios are generated by conservatively extending it. What matters is not simply that reasoning go beyond what is explicitly given, but that it do so while preserving the common core and determining only content that had not already been established.

From this perspective, Default Logic comes close because it also moves from a base to a plurality of further outcomes. Yet it drives those outcomes through defeasible inference by means of default rules: it provides admissible conclusions in the absence of defeating information, rather than the conservative determination of an open region within a shared partial factual base. Its determinations therefore arise from rule-governed nonmonotonic inference rather than from conjectural extensions beyond settled facts.

Autoepistemic Logic has as its center of gravity belief reflection: the relevant operator tracks what an agent believes or fails to believe, and the resulting stable expansions are shaped by introspective doxastic commitments. Conjectures, on the other hand, do not begin from self-ascription of belief, but from a shared factual core and the controlled production of conjectural scenarios extending that core only where it remains unsettled.

AGM Belief Revision also concerns the transition from one state of commitment to another, and in that respect it naturally invites comparison. Still, the operative phenomenon is rational theory change under incoming information: expansion, contraction, and revision describe how an accepted belief set should be modified when new information must be incorporated. Conjectures are not concerned with the rational reorganization of a theory under informational pressure, but with the extension of an incomplete base through admissible additions that preserve what is already settled in it. This difference is also relevant for the discussion about the extension to dynamic modal logic we propose: the operator \textit{settle} should not be read either as AGM-style revision or as a standard update mechanism in the sense of traditional Dynamic Epistemic Logic, as it does not repair a belief set nor transmit new information through an epistemic model, but aims at increasing the determination of a previously open factual core.

\textit{Ceteris paribus} logics also come close to the present framework in one important respect: they make explicit the idea that reasoning may proceed by varying some elements of a situation while preserving a stable background. This resembles the role played here by the shared core of a partial factual base. Yet the similarity stops at that level. In ceteris paribus reasoning, the main task is to compare situations under a constraint of contextual equality and to determine which differences matter once the rest is held unchanged. In conjectural reasoning, instead, the main task is to continue a shared but incomplete factual base by determining content that had not yet been settled and then exploring the resulting scenario.

The expressive role of the two frameworks is therefore different. Ceteris paribus logics are primarily comparative: they articulate constrained variation among situations already available for evaluation. Conjectures are instead generative: they introduce a new cognitive commitment that prolongs a partial base without revising its settled core. For this reason, ceteris paribus logics illuminate one structural aspect of conjectures, namely the preservation of a common background, but they do not provide a direct characterization of conjecture as it is understood in this paper.

Standpoint Logic perhaps comes closest to conjectures when considering the corresponding levels of plurality. It allows different perspectives to organize the same domain differently and gives formal space to viewpoint-dependent commitments. Yet that feature remains much broader than what is described here by conjectures. Conjectural scenarios do not arise from globally different standpoints over the same domain, but from a shared settled core that is taken as binding and varied only where that core leaves room open. The plurality relevant to conjecture is therefore narrower and more structured than the plurality of standpoints or acceptable precisifications.

What unifies the peculiarity of conjectures is the role of the base: here, conjectures are neither a defeasible conclusion, nor a reflective belief state, nor a revised theory, nor a perspective taken as a whole, but a cognitive commitment generated by conservatively extending a shared partial factual base in a non-bivalent setting. This requirement of sharedness, partiality, and preservation is precisely what the neighboring frameworks illuminate without fully capturing.

None of these four frameworks is therefore an acceptable substitute for KC and KDC. Each models a nearby phenomenon with genuine logical interest, but none of them captures conjecture in a way sufficiently close to the notion developed in this paper. Only a framework based on non-bivalent partial factual bases and conservative extensions gives conjecture the intended cognitive structure.

\section{An Alethic Reading of Axiom \textbf{C}}\label{alethicReadingOfC}

The hypotheses and analysis developed in this paper were primarily motivated by a cognitive or informational interpretation of modal logics, and the introduction of Axiom \textbf{C} was meant to extend cognitive logics beyond doxastic and epistemic logics, towards a direction that we felt was underexplored. In a cognitive or informational setting, the accessibility relation of modal logics is naturally understood in terms of informational extension, and \textbf{C} expresses a form of preservation of established truths across informationally compatible worlds. This is how we would like the model introduced here to be considered and evaluated. 

Nonetheless, we feel tempted to entertain also a different interpretation of Axiom \textbf{C} in a non-bivalent setting, based on an alethic understanding of the modal operator and ignoring for the moment cognitive interpretations. 

In a classical and bivalent reading, the valuation of non-modal statements is already established in full: either something is a fact or its negation is. The modal operator $\Box$ identifies what is \textit{necessary} to happen, because it is happening in all worlds accessible to the current one, while the $\Diamond$ operator allows us to characterize \textit{possibility} as something that could have happened or not, regardless of whether it eventually did, since there are accessible worlds in which it holds and others in which it does not: it is a counterfactual possibility of things that in the current world have already turned out in one way. 

Yet, when the underlying semantics is non-bivalent, a different reading is appropriate. In a non-bivalent setting, not all facts have been established in full: statements may be undetermined at the current world, and thus neither they nor their negation are facts. 

Within such a framework, the modal structure allows a natural alethic interpretation. Undetermined statements are eventual outcomes of an open-ended evaluation that has not (or cannot) be performed: they represent not \textit{counterfactual possibility} ("$\varphi$ is true but could have been false"), but \textit{undetermined possibility} ("$\varphi$ is neither true nor false, and it can still become one or the other"). Philosophically, this reading is indifferent to whether the openness is the result of informational incompleteness in a fully deterministic universe, or of a random outcome of a genuinely non-deterministic universe. 

Axiom \textbf{C} expresses therefore a principle of \textit{stability} of determined facts: if a proposition is true at the current world, then it holds in all accessible worlds. Its intuitive meaning becomes immediate and clear: facts remain facts when undetermined propositions become determined, and once a proposition has become a fact, it cannot and will not be retracted.

\section{Conclusions}\label{sec:conclusions}

In this paper we have introduced \textit{conjectural logics}, a new family of cognitive modal systems that complement well-known doxastic and epistemic frameworks. These logics are grounded on a non-bivalent, non-reflexive formulation of the schema
$\varphi \rightarrow \Box \varphi$ (Axiom \textbf{C}), enabling a principled distinction between factual information and conjectural reasoning. We have shown that Axiom \textbf{C} does not by itself lead to modal collapse, which arises only in the presence of Axiom \textbf{T} or under concretely bivalent assumptions on the base logic. On this basis, we formally defined the systems \textbf{KC} and \textbf{KDC}, showed that \textbf{C} trivially implies \textbf{4} and \textbf{5}, and proved that these systems are non-trivial, sound, and complete.

This approach, relying on set-theoretic inclusion relations between valuations in designated worlds, gives rise to a clear and general inclusion theorem, allowing us to interpret different cognitive modes, such as doxastic, epistemic, conjectural, delusional, and opinion-based, as relationships between partial valuation sets rather than as wholly separate syntactic systems.

Doxastic, epistemic, and conjectural worlds all admit a simple constructive semantics for modelling partial, potentially incorrect, or reality-disconnected cognitive states. In this way the framework supports the representation of uncertainty, disagreement, and conflicting opinions within a unified setting, and captures credence, delusion, subjective stance, and conjectural extension in formally comparable terms.

The framework also admits a dynamic development. By introducing the operator $\mathsf{settle}(p)$, we showed how a conjectural extension may become the new designated reality, thereby motivating a corresponding \textit{Conjectural Dynamic Logic}. This dynamic layer makes it possible to represent not only static conjectural states, but also transitions by which an incomplete factual base becomes progressively more determined.

Beyond its theoretical interest, this model has significant potential for practical applications in fields such as the Semantic Web, knowledge provenance, and neuro-symbolic AI. For example, in provenance-aware systems researchers have studied how to represent and reason about conflicting claims and their sources, using structured provenance models to track the credibility of assertions \cite{sikos2020provenance} \cite{leech_provenance2023}. Similarly, in neuro-symbolic AI, hybrid architectures that combine neural learning with formal symbolic logic offer a way to enrich systems with both interpretability and robust reasoning capabilities \cite{hitzler2022neurosymbolic}. The conjectural framework aligns with these hybrid systems: agents or modules may share partial knowledge bases, make conjectures based on known facts, and update them dynamically as new information becomes definite, mirroring both the semantics of settlement events and the architecture of explainable neuro-symbolic reasoning.

Overall, we propose a unified, formally grounded cognitive logic that bridges the expressive gap between belief, knowledge, and conjecture, handles partial and undefined information without collapse, supports dynamic transitions from conjectural extensions to designated reality via settlement operations, and integrates naturally with emerging technologies in symbolic and neuro-symbolic AI.

Future work will explore computational implementations of algorithmic reasoning for conjectural systems, and concrete case studies in multi-agent settings, provenance tracking, and neuro-symbolic integration.

\section*{Acknowledgments}\label{sec:ack}
TBD

\bibliography{Content}

@incollection{anderson1996,
  title={G{\"o}del's ontological proof revisited},
  author={Anderson, C. Anthony and Gettings, Michael},
  booktitle={G{\"o}del'96: Logical foundations of mathematics, computer science and physics---Kurt G{\"o}del's legacy, Brno, Czech Republic, August 1996, proceedings},
  volume={6},
  pages={167--173},
  year={1996},
  publisher={Association for Symbolic Logic}
}

@article{benzmueller2025scott,
  author  = {Christoph Benzmüller and Dana Scott},
  title   = {Notes on {Gödel's and Scott's} Variants of the Ontological Argument},
  journal = {Monatshefte für Mathematik},
  year    = {2025},
  doi     = {10.1007/s00605-025-02078-x},
  note    = {Includes full transcription of Kurt Gödel’s unpublished ontological proof from his Princeton Nachlass.}
}

@book{blackburn2001modal,
  title     = {Modal Logic},
  author    = {Blackburn, Patrick and de Rijke, Maarten and Venema, Yde},
  publisher = {Cambridge University Press},
  year      = {2001},
  isbn      = {9780521527146}
}

@book{chagrov1997modal,
  author    = {Alexander Chagrov and Michael Zakharyaschev},
  title     = {Modal Logic},
  publisher = {Oxford University Press},
  year      = {1997}
}

@phdthesis{fervari2014relation,
  title = {Relation-changing modal logics},
  author = {Fervari, R\'emi},
  school = {Universidad Nacional de Córdoba},
  year = {2014},
  url = {https://core.ac.uk/download/pdf/344694710.pdf}
}

@article{girard2019modal,
  title = {Modal logic without contraction in a metatheory without contraction},
  author = {Girard, Patrick and Weber, Zach},
  journal = {The Review of Symbolic Logic},
  volume = {12},
  number = {4},
  pages = {707--732},
  year = {2019},
  doi = {10.1017/S1755020318000292},
  url = {https://www.cambridge.org/core/journals/review-of-symbolic-logic/article/modal-logic-without-contraction-in-a-metatheory-without-contraction/1F712CD8DA333AAED7D77C4677713955}
}

@misc{godel1970ontological,
  author = {Kurt Gödel},
  title  = {Ontological Proof Manuscript},
  year   = {ca. 1970},
  note   = {Unpublished manuscript written circa 1970, found in Gödel's Nachlass, Institute for Advanced Study, Princeton. Full transcription in: Benzmüller \& Scott (2025).}
}

@article{hitzler2022neurosymbolic,
  author    = {Pascal Hitzler and Aaron Eberhart and Monireh Ebrahimi and Md Kamruzzaman Sarker and Lu Zhou},
  title     = {Neuro-symbolic approaches in artificial intelligence},
  journal   = {National Science Review},
  volume    = {9},
  number    = {6},
  pages     = {nwac035},
  year      = {2022},
  doi       = {10.1093/nsr/nwac035}
}

@incollection{kovac2012modal,
  author    = {Sre{\'c}ko Kova{\v{c}}},
  title     = {Modal collapse in {G{\"o}del's} ontological proof},
  booktitle = {Ontological Proofs Today},
  editor    = {Miroslaw Szatkowski},
  publisher = {Ontos Verlag},
  address   = {Heusenstamm},
  year      = {2012},
  pages     = {323--344},
  url       = {https://urn.nsk.hr/urn:nbn:hr:261:299546}
}

@article{leech_provenance2023,
  title        = {Provenance-Based Trust-Aware Requirements Engineering Framework for Self-Adaptive Systems},
  author       = {Lee, Hyo‑Cheol and Lee, Seok‑Won},
  journal      = {Sensors},
  volume       = {23},
  number       = {10},
  pages        = {4622},
  year         = {2023},
  doi          = {10.3390/s23104622}
}

@article{lewitzka2017amalgam,
  title = {A modal logic amalgam of classical and intuitionistic propositional logic},
  author = {Lewitzka, Sigrid},
  journal = {Journal of Logic and Computation},
  volume = {27},
  number = {1},
  pages = {201--222},
  year = {2017},
  doi = {10.1093/logcom/exw018},
  url = {https://arxiv.org/pdf/1306.2068}
}

@mastersthesis{liao2023intuitionistic,
  author    = {Cheng Liao},
  title     = {Stable Canonical Rules for Intuitionistic Modal Logics},
  school    = {Universiteit van Amsterdam},
  type      = {Master's thesis},
  year      = {2023},
  url       = {https://eprints.illc.uva.nl/id/eprint/2248/1/MoL-2023-07.text.pdf},
  note      = {Institute for Logic, Language and Computation (ILLC)}
}

@article{marcos2005paranormal,
  title     = {Nearly every normal modal logic is paranormal},
  author    = {Marcos, Jo{\~a}o},
  journal   = {Logique et Analyse},
  volume    = {48},
  number    = {189--192},
  pages     = {279--300},
  year      = {2005}
}

@book{oppy1995ontological,
  author    = {Graham Oppy},
  title     = {Ontological Arguments and Belief in God},
  publisher = {Cambridge University Press},
  year      = {1995}
}

@article{sikos2020provenance,
  title = {Provenance‑Aware Knowledge Representation: A Survey of Data Models and Contextualized Knowledge Graphs},
  author = {Sikos, Leslie F. and Philp, Dean},
  journal = {Data Science and Engineering},
  volume = {5},
  number = {3},
  pages = {293--316},
  year = {2020}
}

@book{sobel2004logic,
  author    = {Jordan Howard Sobel},
  title     = {Logic and Theism: Arguments for and against Beliefs in God},
  publisher = {Cambridge University Press},
  year      = {2004}
}

@Article{solovay1976,
  author  = {Robert M. Solovay},
  title   = {Provability Interpretations of Modal Logic},
  journal = {Israel Journal of Mathematics},
  volume  = {25},
  pages   = {287--304},
  year    = {1976}
}

@article{vanbenthem2007dynamic,
author = {Johan van Benthem},
title = {Dynamic logic for belief revision},
journal = {Journal of Applied Non-Classical Logics},
volume = {17},
number = {2},
pages = {129--155},
year = {2007},
publisher = {Taylor \& Francis},
doi = {10.3166/jancl.17.129-155},
}

@article{drobyshevich2020,
  author  = {Sergey Drobyshevich and Heinrich Wansing},
  title   = {Proof Systems for Various FDE-Based Modal Logics},
  journal = {The Review of Symbolic Logic},
  year    = {2020},
  volume  = {13},
  number  = {4},
  pages   = {720--747},
  doi     = {10.1017/S1755020319000261}
}

@article{DalmonteGrelloisOlivetti2020,
  author  = {Tiziano Dalmonte and Charles Grellois and Nicola Olivetti},
  title   = {Intuitionistic Non-normal Modal Logics: A General Framework},
  journal = {Journal of Philosophical Logic},
  year    = {2020},
  volume  = {49},
  pages   = {833--882},
  doi     = {10.1007/s10992-019-09539-3}
}

@article{figallo2022,
  author  = {Aldo Figallo{-}Orellano and Miguel P{\'e}rez{-}Gaspar and Juan Manuel Ram{\'\i}rez{-}Contreras},
  title   = {Paraconsistent and Paracomplete Logics Based on k{-}Cyclic Modal Pseudocomplemented {De Morgan} Algebras},
  journal = {Studia Logica},
  year    = {2022},
  volume  = {110},
  number  = {5},
  pages   = {1291--1325},
  doi     = {10.1007/s11225-022-10004-7}
}

@article{Hodes2021Part1,
  author  = {Harold T. Hodes},
  title   = {One-Step Modal Logics, Intuitionistic and Classical, Part 1},
  journal = {Journal of Philosophical Logic},
  year    = {2021},
  volume  = {50},
  number  = {5},
  pages   = {837--872},
  doi     = {10.1007/s10992-020-09574-5}
}

@article{pawlowski2024,
  author  = {Pawel Pawlowski and Daniel Skurt},
  title   = {8 Valued Non-Deterministic Semantics for Modal Logics},
  journal = {Journal of Philosophical Logic},
  year    = {2024},
  volume  = {53},
  number  = {2},
  pages   = {351--371},
  doi     = {10.1007/s10992-023-09733-4}
}

@article{drucker2011factaCapta,
	author = {Drucker, Johanna},
	title = {Humanities Approaches to Graphical Display},
	journal = {Digital Humanities Quarterly},
	year = {2011},
	volume = {5},
	number = {1},
	url = {https://dhq-static.digitalhumanities.org/pdf/000091.pdf}
}

@article{Reiter1980Default,
  author  = {Raymond Reiter},
  title   = {A Logic for Default Reasoning},
  journal = {Artificial Intelligence},
  volume  = {13},
  number  = {1--2},
  pages   = {81--132},
  year    = {1980},
  doi     = {10.1016/0004-3702(80)90014-4}
}

@article{Moore1985Autoepistemic,
  author  = {Robert C. Moore},
  title   = {Semantical Considerations on Nonmonotonic Logic},
  journal = {Artificial Intelligence},
  volume  = {25},
  number  = {1},
  pages   = {75--94},
  year    = {1985},
  doi     = {10.1016/0004-3702(85)90042-6}
}

@article{AlchourronGardernforsMakinson1985AGM,
  author  = {Carlos E. Alchourr{\'o}n and Peter G{\"a}rdenfors and David Makinson},
  title   = {On the Logic of Theory Change: Partial Meet Contraction and Revision Functions},
  journal = {The Journal of Symbolic Logic},
  volume  = {50},
  number  = {2},
  pages   = {510--530},
  year    = {1985},
  doi     = {10.2307/2274239}
}

@incollection{GomezAlvarezRudolph2022Standpoint,
  author    = {Luc{\'\i}a G{\'o}mez {\'A}lvarez and Sebastian Rudolph},
  title     = {Standpoint Logic: Multi-Perspective Knowledge Representation},
  booktitle = {Formal Ontology in Information Systems},
  editor    = {F. Neuhaus and B. Brodaric},
  publisher = {IOS Press},
  year      = {2022},
  pages     = {3--17},
  doi       = {10.3233/FAIA210367}
}

@article{vanBenthemGirardRoy2009CeterisParibus,
  author  = {Johan van Benthem and Patrick Girard and Olivier Roy},
  title   = {Everything Else Being Equal: A Modal Logic for Ceteris Paribus Preferences},
  journal = {Journal of Philosophical Logic},
  volume  = {38},
  number  = {1},
  pages   = {83--125},
  year    = {2009},
  doi     = {10.1007/s10992-008-9085-3}
}

@inproceedings{Plaza1989PublicAnnouncements,
	author = {Plaza, Jan A.},
	title = {Logics of Public Communications},
	booktitle = {Proceedings of the 4th International Symposium on Methodologies for Intelligent Systems},
	pages = {201--216},
	year = {1989},
	publisher = {North-Holland}
}

@article{Fine1975Supervaluation,
	author = {Fine, Kit},
	title = {Vagueness, Truth and Logic},
	journal = {Synthese},
	volume = {30},
	number = {3--4},
	pages = {265--300},
	year = {1975},
	doi = {10.1007/BF00485047}
}

@book{Kleene1952Metamathematics,
	author = {Kleene, Stephen Cole},
	title = {Introduction to Metamathematics},
	publisher = {North-Holland},
	address = {Amsterdam},
	year = {1952}
}

@book{BaaderEtAl2003DLHandbook,
	editor = {Baader, Franz and Calvanese, Diego and McGuinness, Deborah L. and Nardi, Daniele and Patel-Schneider, Peter F.},
	title = {The Description Logic Handbook: Theory, Implementation, and Applications},
	publisher = {Cambridge University Press},
	address = {Cambridge},
	year = {2003}
}

@article{Chaudhuri2023BRCA1Controversy,
	author = {Chaudhuri, A. R. and Nussenzweig, A.},
	title = {The ubiquitin ligase activity of {BRCA1--BARD1} and its role in DNA repair and tumor suppression},
	journal = {Molecular Cell},
	volume = {83},
	number = {20},
	pages = {3777--3780},
	year = {2023},
	doi = {10.1016/j.molcel.2023.09.019}
}

@article{Findlay2023BRCA1Dispensable,
  author = {Findlay, Suzanne and Heath, Jennifer and Luo, Vicky M. and others},
  title = {BRCA1--BARD1 E3 ligase activity is dispensable for homology-directed DNA repair and for tumour suppression},
  journal = {The Journal of Pathology},
  year = {2023},
  volume = {261},
  number = {4},
  pages = {507--523},
  doi = {10.1002/path.6199}
}

\end{document}